\documentclass[times, twoside]{zHenriquesLab-StyleBioRxiv}
\usepackage[left]{lineno}
\leadauthor{Tanaka} 

\begin{document}

\hyphenpenalty=2000\relax
\exhyphenpenalty=2000\relax
\sloppy

\title{\LARGE Dynamic change of gene-to-gene regulatory networks in response to SARS-CoV-2 infection}
\shorttitle{SARS-CoV-2 network analysis}

\author[1,2]{Yoshihisa Tanaka}
\author[3]{Kako Higashihara}
\author[3]{Mai Adachi Nakazawa}
\author[1]{Fumiyoshi Yamashita}
\author[3*\Letter]{Yoshinori Tamada}
\author[2,3*\Letter]{\\Yasushi Okuno}

\affil[1]{Graduate School of Pharmaceutical Sciences, Kyoto University}
\affil[2]{RIKEN Cluster for Science, Technology and Innovation Hub, Medical Sciences Innovation Hub Program}
\affil[3]{Graduate School of Medicine, Kyoto University}

\maketitle

\begin{abstract}
\section*{Abstract}
The current pandemic of SARS-CoV-2 has caused extensive damage to society. The characterization of SARS-CoV-2 prpfiles has been addressed by researchers globally with the aim of resolving this disruptive crisis. This investigation process is indispensable for an understanding of how SARS-CoV-2 behaves in human host cells. However, little is known about the systematic molecular mechanisms involved in the effect of SARS-CoV-2 infection on human host cells. Here, we have presented gene-to-gene regulatory networks in response to SARS-CoV-2 using a Bayesian network model. We examined the dynamic changes of the SARS-CoV-2-purturbated networks established by our proposed framework for gene network analysis, revealing that interferon signaling gradually switches to the subsequent inflammatory-cytokine signaling cascades. Furthermore, we have succeeded in capturing a COVID-19 patient-specific network in which transduction of these signalings is coincidently induced. This enabled us to explore local regulatory systems influenced by SARS-CoV-2 in host cells more precisely at an individual level. Our panel of network analyses has provided new insight into SARS-CoV-2 research from the perspective of cellular systems.
\end {abstract}

\begin{keywords}
SARS-CoV-2 | gene network analysis | COVID-19 individual network
\end{keywords}

\begin{corrauthor}
tamada.yoshinori.8a@kyoto-u.ac.jp,\\okuno.yasushi.4c@kyoto-u.ac.jp
\end{corrauthor}

\section*{Introduction}

\begin{figure*}[ht]
\begin{center}
\includegraphics[width=.8\linewidth]{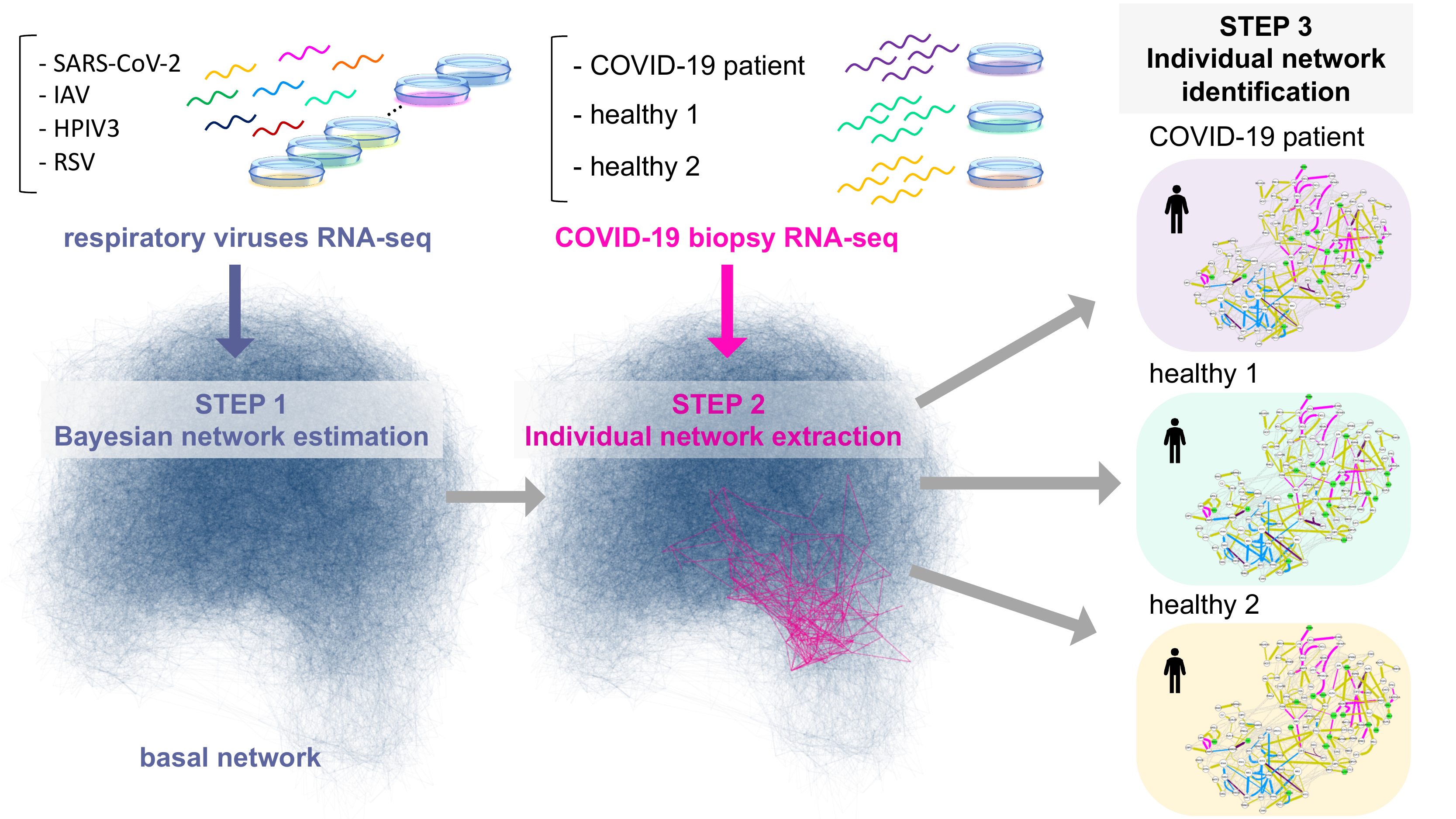}
\end{center}
\caption{Illustration of overview.
The hairball (blue) is the basal network consisting of 127,126 edges and 15,258 nodes established by use of the respiratory viruses RNA-seq including SARS-CoV-2. The highlighted-network (magenta) in the basal network represents the COVID-19-perturbated network extracted by using the biopsy RNA-seq.}
\label{fig1}
\end{figure*}

The newly emerging coronavirus, severe acute respiratory syndrome coronavirus 2 (SARS-CoV-2), has spread rapidly over the world \cite{Zhu2020-wm,Wu2020-tu}, with more than 22,000,000 cases of coronavirus disease 2019 (COVID-19) and 790,000 deaths as of August 21, 2020 \cite{Dong2020-xn}. This pandemic outbreak has drastically changed our society, and has compelled us to stay alert to the continuous risk of SARS-CoV-2 infection \cite{Hellewell2020-np}. To overcome this dire situation, the development of novel drugs or vaccines is still an urgent global challenge. During the therapeutic development process, the elucidation of cellular mechanisms is essential for the discovery of potential targets; the fundamental question to be solved is how SARS-CoV-2 influences host cells and causes COVID-19 at the molecular level. However, these cellular mechanisms for COVID-19 are poorly understood.

High-throughput technologies have contributed to the acquisition of a large amount of “omics” data, which has provided comprehensive information on the cellular systems. These technologies have also featured during the current research into SARS-CoV-2. Several reports have provided various clues to understanding the global cellular signatures in response to SARS-CoV-2 infection at both the proteome and transcriptome level \cite{Gordon2020-sf,Bojkova2020-hu,Blanco-Melo2020-cv}. Recently, network-based approaches have stimulated great interest in the use of these emerging omics data for drug discovery and systems biological analysis in the current SARS-CoV-2 field \cite{Yan2018-hu,Recanatini2020-jo,Guzzi2020-rr,Zhou2020-uf,Gysi2020-pd,Fagone2020-nz}. Their major approaches combine publicly available sources, including knowledge of the already established pathways and drugs with these omics data to reconstruct molecular networks. However, these networks do not sufficiently represent a real-world cellular system owing to two main reasons: 1) public data consist of heterogeneous knowledge that has been accumulated throughout the longstanding biological researches; and 2) the previous works use mixed networks that combine data from various samples, but cannot reflect an individual cell-/patient- specific network.

To address these problems, we recently developed a method to extract a core sample-specific network from a massive gene network generated from a Bayesian network \cite{Tanaka2020-xr}. Gene regulatory network estimation has been developed as a prospective method to model the cellular system using omics data \cite{Margolin2006-oc,Araki2009-lf,Wang2012-qb,Affara2013-vc,Singh2018-sz}. Although Bayesian network-based approach can infer the cause-and-effect relationships between genes with transcriptome data, the key issue has been to extract biologically significant information from the huge and complicated network, which is often sarcastically referred to as a hairball \cite{Yan2016-hu}. Our unique framework consists of three steps: 1) estimation of a global gene network; 2) extraction of context-specific core networks based on differences in molecular systems from the global network; and 3) identification of a sample or patient specific network (Fig.~\ref{fig1}). The prominent advantage is that it enables us to identify putative context-specific or sample-specific potential sets of edges in the form of a network, i.e., gene-to-gene relationships with directions, as well as nodes. 

In this study, by using our developed framework for gene network analysis, we have presented the core host cellular systems involved in SARS-CoV-2 over several {\it in vitro} experiments; different viral loads, cell lines, and respiratory viruses. No studies have been performed on the computational data-driven gene regulatory network approach regarding SARS-CoV-2. We characterized interferon signaling and subsequent inflammatory signaling cascades as significantly changed networks in human host cells, which represent the innate antiviruses-immune system in response to SARS-CoV-2 infection. In addition, given that the recent studies have reported that patients with COVID-19 exhibit various clinical outcomes depending on each patient, and that a certain proportion of patients will experience a severe disease \cite{Huang2020-vp,Guan2020-ci,Mehta2020-fc}, it is much more important to reveal the cellular mechanisms causing these clinical symptoms at an individual level. To this end, we have further identified the gene networks specifically for patients with COVID-19. We believe that our landscape of gene networks is beneficial to achieve an understanding of how cellular systems respond to SARS-CoV-2 and to further drug development.

\section*{Results}

\subsection*{Estimation of basal gene network in the involvement of respiratory viruses infection using a Bayesian network}

We first characterized a global gene network (hereafter referred to as the {\it basal network}) using a Bayesian network (see Methods) with a transcriptome dataset involved in the engagement of respiratory virus infection, including SARS-CoV-2, in several human cell lines \cite{Blanco-Melo2020-cv}. To determine a basal network structure, we performed a network estimation using the neighbor node sampling and repeat algorithm \cite{Tamada2011-ub}, and screened the best algorithm parameters for the target dataset, as described in our previous study \cite{Tanaka2020-xr}. Briefly, the network estimation was run three times independently, and the subsequent concordance test was performed to ensure the robustness and stability of the estimated basal network. We confirmed that the iteration number $T=500,000$ satisfies less than 5$\%$ error (Error = 4.0$\%$ for $T=500,000$;  error = 5.3$\%$ for $T=300,000$). The final basal network comprised 127,126 edges and 15,258 nodes, with a threshold of 0.05 and an average degree of 16.7. We used this final basal network for the subsequent analyses.

\subsection*{Dynamics of host cellular network profiles in different viral loads of SARS-CoV-2}

\begin{figure*}[ht]
\begin{center}
\includegraphics[width=.8\linewidth]{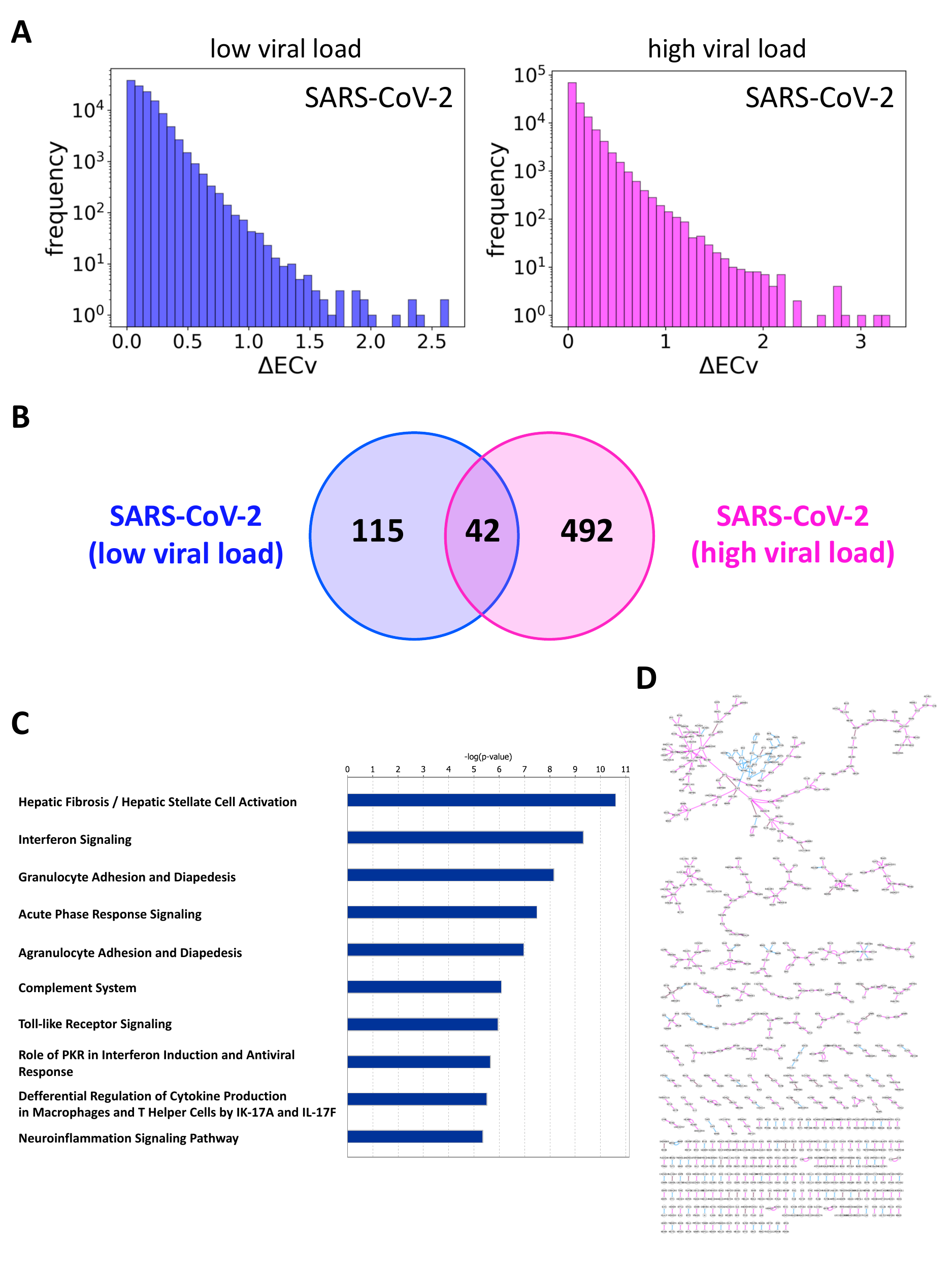}
\end{center}
\caption{Dynamics of the SARS-CoV-2-perturbated network for different viral loads in host cells.
(\textbf{A}) The histograms of $\Delta$ECv for different SARS-CoV-2 viral loads; a low MOI of 0.2 (blue) and a high MOI of 2 (magenta). The X-axis corresponds to the threshold for each $\Delta$ECv. The Y-axis shows the number of edges on a log scale. (\textbf{B}) The Venn diagram represents the numbers of differentially regulated edges (DREs) for two SARS-CoV-2 viral loads (blue: low MOI, magenta: high MOI) with a threshold of 1.0 for $\Delta$ECv in Fig. 2A. (\textbf{C}) The top 10 terms of canonical pathway analysis for the genes comprising a union set of $\Delta$ECv-extracted DREs in the Venn diagram analysis (Fig. 2B). (\textbf{D}) The whole picture for the various sizes of subnetwork fragments is shown. Image of how the $\Delta$ECv-extracted DREs mutually connected and generated the subnetworks.}
\label{fig2}
\end{figure*}

To examine the transition of host cellular system dynamics during the increase of SARS-CoV-2 viral loads, we aimed to characterize the networks perturbated by SARS-CoV-2 with two viral loads; a low multiplicity of infection (MOI) of 0.2 and a high MOI of 2 in A549 cells. We expected that cells exposed to different viral loads each present a unique cellular system, and that our approach could capture the fluctuation of system dynamics in whole cellular systems. To obtain differential core gene networks for each viral load, we followed the multiple steps using an edge quantification technique, called {\it edge contribution value} (ECv), established in our previous study \cite{Tanaka2020-xr}. We first calculated $\Delta$ECvs following the Eq.~(\ref{eq:ecv}) where $S$ = SARS-CoV-2 infected and $T$ = mock samples for each MOI condition (see Methods). The distributions of $\Delta$ECv showed that the innate cellular system was more extensively perturbated in the cells exposed to the high MOI than the low MOI  (\textbf{Fig.~\ref{fig2}A}). We next set a threshold of 1 for $\Delta$ECv and obtained {\it differentially regulated edges} (DREs) from the basal network. The Venn analysis for the $\Delta$ECv-extracted DREs showed that the number of DREs in the high MOI was larger than in the low MOI (\textbf{Fig.~\ref{fig2}B}). Interestingly, the number of shared DREs between high- and low- MOIs was 42, which was only 6$\%$ of the total number of DREs in both conditions, indicating that the underlying regulatory system between them was not similar. To confirm the biological involvement of the DREs, we performed canonical pathway analysis for the genes contained in the $\Delta$ECv-obtained DREs, showing that these genes were associated with some cellular antiviral systems (\textbf{Fig.~\ref{fig2}C}). These results supported that the components of the DREs were biologically relevant to viral infection.

\begin{figure*}[t]
\begin{center}
\includegraphics[width=\linewidth]{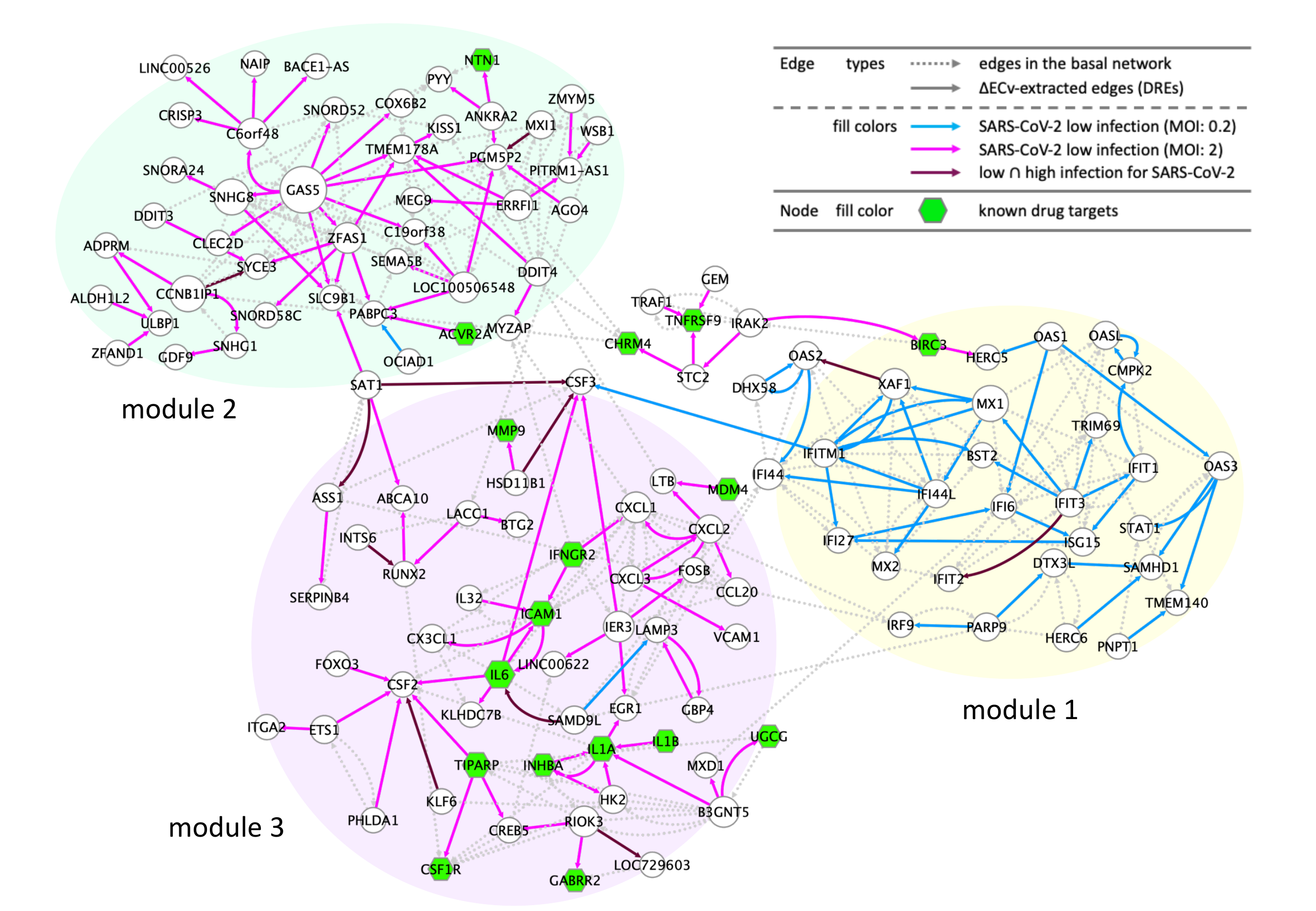}
\end{center}
\caption{The SARS-CoV-2-perturbated host cellular network in response to different viral loads in A549 cells (the SARS-CoV-2-perturbated network). The network comprises 130 nodes and 305 edges (including 155 basal edges). The colored solid edges represent the SARS-CoV-2-perturbated DREs; high MOI of 2 (magenta), low MOI of 0.2 (blue), high MOI $\cap$ low MOI (purple). Dot edges represent the basal edges (grey). The nodes (green) represents the known drug target genes (Supplementary Table S1). The node size stands for the extent of outdegree.}
\label{fig3}
\end{figure*}

To gain a greater insight into the profiles of the DREs from the perspective of network topology, we next generated networks using a set of all the DREs in the Venn diagram (\textbf{Fig.~\ref{fig2}B}). These DREs connected mutually and, in turn, generated various sizes of subnetwork fragments (\textbf{Fig.~\ref{fig2}D}). We reasoned that if these fragments represented some biological significance, these features should be reflected as modularity, as biologically close functions in cellular systems link together and shape modules \cite{Barabasi2004-zv}. Hence, small-sized fragments were likely to be less informative, and we focused on the biggest connected component among the various fragments. The biggest connected component was extracted and the basal edges were additionally mapped on this network, which established the SARS-CoV-2-perturbated network with 130 nodes and 305 edges (\textbf{Fig.~\ref{fig3}}). We found that this network consisted of clear three modules linking each other. One module (module 1, yellow marked region) was mainly formed of a set of the DREs in the low MOI condition, and its constituent elements were many interferon (IFN)-stimulated genes (ISGs): IFIs, MXs, OASs, TRIMs, IFTMs, IRFs, and STATs. These highly orchestrated webs of various ISGs are induced by transductions of both IFN signaling and subsequent JAK/STAT signaling \cite{Mesev2019-za}. This evidence strongly suggested that module 1 represents the consequences of activation of both these signaling pathways by the acute antiviral response. Contrary to module 1, the other two modules (module 2, green marked region; and module 3, purple marked region) are mainly shaped by a set of the DREs in the high MOI condition. Module 2 and module 3 were found to comprise fewer IFN-related genes. While module 2 appeared to be a GAS5-centralized module, module 3 was composed of chemokines (CXCL1, CXCL2, CXCL3, CX3CL1, and CCL20), interleukins (IL6, IL1A, IL1B, and IL32), and colony-stimulating factors (CSF2 and CSF3), which are implicated in inflammatory-related cytokine signaling followed by the acute activation of IFN and JAK/STAT signaling represented in module 1. In particular, the cluster of module 1 and module 3 probably represents the transition of the gene regulatory system in response to SARS-CoV-2 infection. Namely, the cellular system perturbated by SARS-CoV-2 gradually switches to inflammatory signaling (module 3) via IFN and JAK/STAT signaling (module 1) as the viral load increased. This was consistent with the clinical observations of COVID-19, and thus may partially explain the process of cytokine storm syndromes, which is a severe clinical feature of COVID-19 \cite{Huang2020-vp,Mehta2020-fc}. We also performed the same analyses among the four respiratory viruses, and found that module 3 was exclusive for SARS-CoV-2 (\textbf{Supplementary Fig. S1}). Collectively, we identified the SARS-CoV-2-perturbated network and its three modules, which reflected distinctive host cellular functions in response to SARS-CoV-2 infection.

\subsection*{Characterization of the SARS-CoV-2-perturbated network at individual sample level}

\begin{figure*}[t]
\begin{center}
\includegraphics[width=\linewidth]{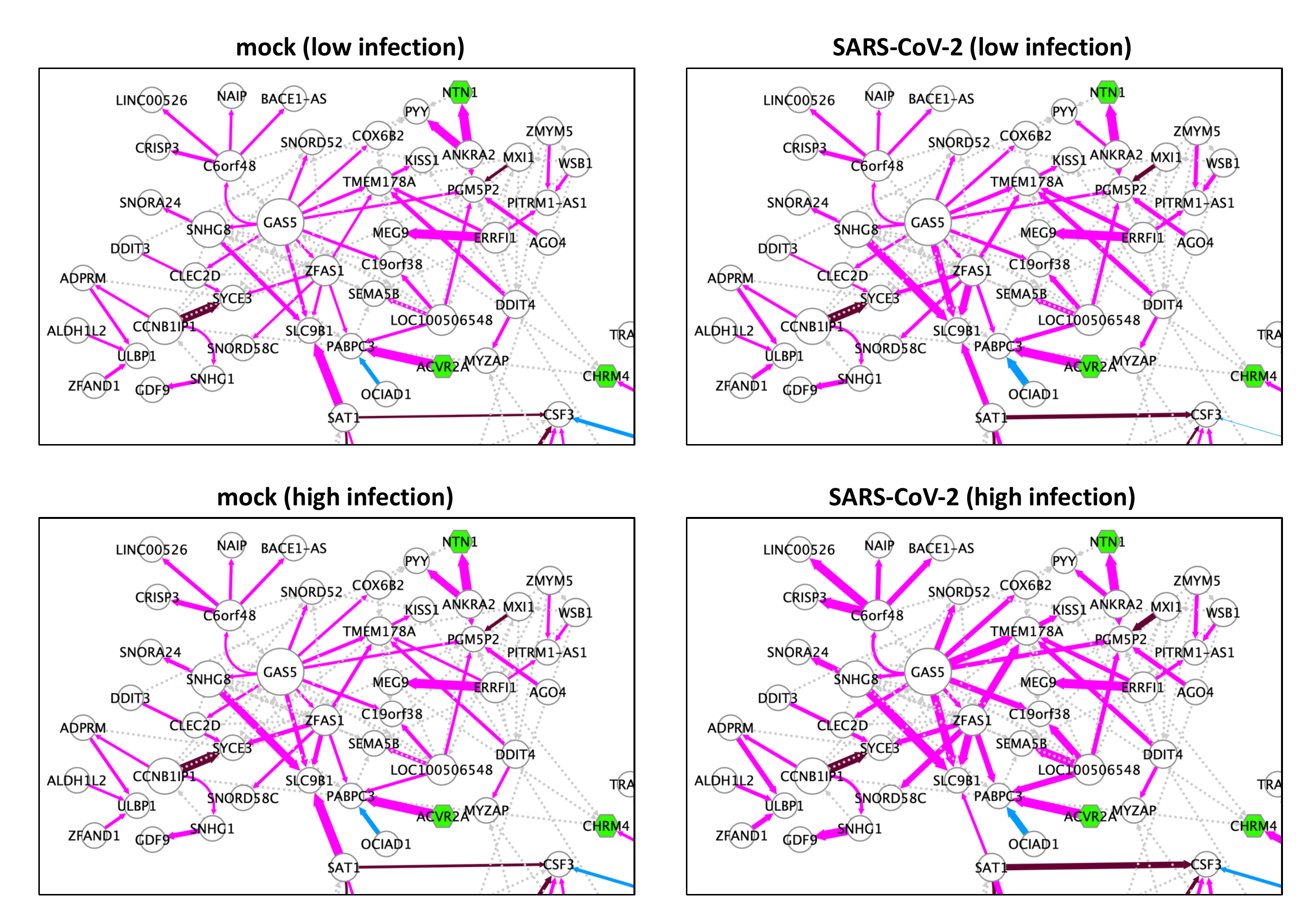}
\end{center}
\caption{Sample-specific individual networks around the GAS5-centralized module.
The GAS5-centralized module (module 3) in the SARS-CoV-2-perturbated network (presented in Fig. 3) is displayed for four representative samples from each group (mock for SARS-CoV-2-infection (low MOI: 0.2), SARS-CoV-2-infected (low MOI: 0.2), mock for SARS-CoV-2-infection (high MOI: 2), and SARS-CoV-2-infected (high MOI: 2)). RCs are represented as edge sizes to show individual differences. The node size stands for the extent of outdegree.
}
\label{fig4}
\end{figure*}

We next determined how the signaling represented in the SARS-CoV-2-perturbated network (\textbf{Fig.~\ref{fig3}}) changed across samples. To this end, we developed a novel quantitative method, called {\it relative contribution} (RC), to measure the edge contribution at an individual level. The mathematical definition of RC is described in Methods. Within a set of pairwise parent-child relations for a certain child, the RC captures how parent genes influence a child gene in response to the pairwise parent’s mRNA expression, and it can therefore reveal local regulatory changes in response to SARS-CoV-2 infection at an individual sample level. To characterize the individual networks, we calculated RCs for 12 samples within four groups (mock $\times$ 3 for SARS-CoV-2-infected (MOI: 0.2), SARS-CoV-2-infected $\times$ 3 (MOI: 0.2), mock $\times$ 3 for SARS-CoV-2-infected (MOI: 2), SARS-CoV-2-infected $\times$ 3 (MOI: 2)) involved in the network generation process in \textbf{Fig.~\ref{fig3}}, and selected representative four samples from each group. By representing RCs as the sizes of edge widths, we depicted these four sample-specific individual networks (\textbf{Supplementary Animation 1}), and found that the vicinity of the GAS5-centralized module (module 2) drastically changed at an RC level (\textbf{Fig.~\ref{fig4}}). Interestingly, this module included GAS5, SNHG8, ZFAS1, SNORD52, SNORD58C, SNORA24, and LOC100506548, which encode non-coding RNA (ncRNA) genes. Given that GAS5 appears to function as a hub gene, these results suggested that the genes downstream of GAS5 are regulated by different cellular systems in the mock and SARS-CoV-2 infections at a local system level. Especially, our results showed that GAS5, ZFAS1, and SNHG8 were found to be dominant for SLC9B1 in SARS-CoV-2-infected samples compared with the mock, suggesting the regulatory system used was significantly different between them (\textbf{Fig.~\ref{fig4}}). GAS5 is a single-strand long ncRNA and one study demonstrated that the mRNA expression of GAS5 was elevated in response to hepatitis C virus infection and that GAS5 impaired virus replication by the interaction between truncated-GAS5 and HCV NS3 protein in human cells \cite{Qian2016-oi}. Combined with this evidence, our results suggest the possibility that this ncRNA-related module 2 may play a novel clear role in SARS-CoV-2 infection. 

Conversely, of the four individual networks, the two networks for mocks exhibited no significant change in RC (\textbf{Fig.~\ref{fig4} and Supplementary Animation 1}). This was consistent with the prerequisite experimental designs as the mock samples are supposed to exhibit the same behaviour, which further supported the validity of our method. Moreover, the RC-highlighted edges displaying even small or no changes explained that their local regulatory system, presented as a set of pairwise parent-child relationships for one child, did not change between the individual samples. Collectively, our data have demonstrated that we can capture the local system differences in network signaling at an individual level.

\subsection*{Identification of specific individual networks for patients with COVID-19}

\begin{figure*}[t]
\begin{center}
\includegraphics[width=\linewidth]{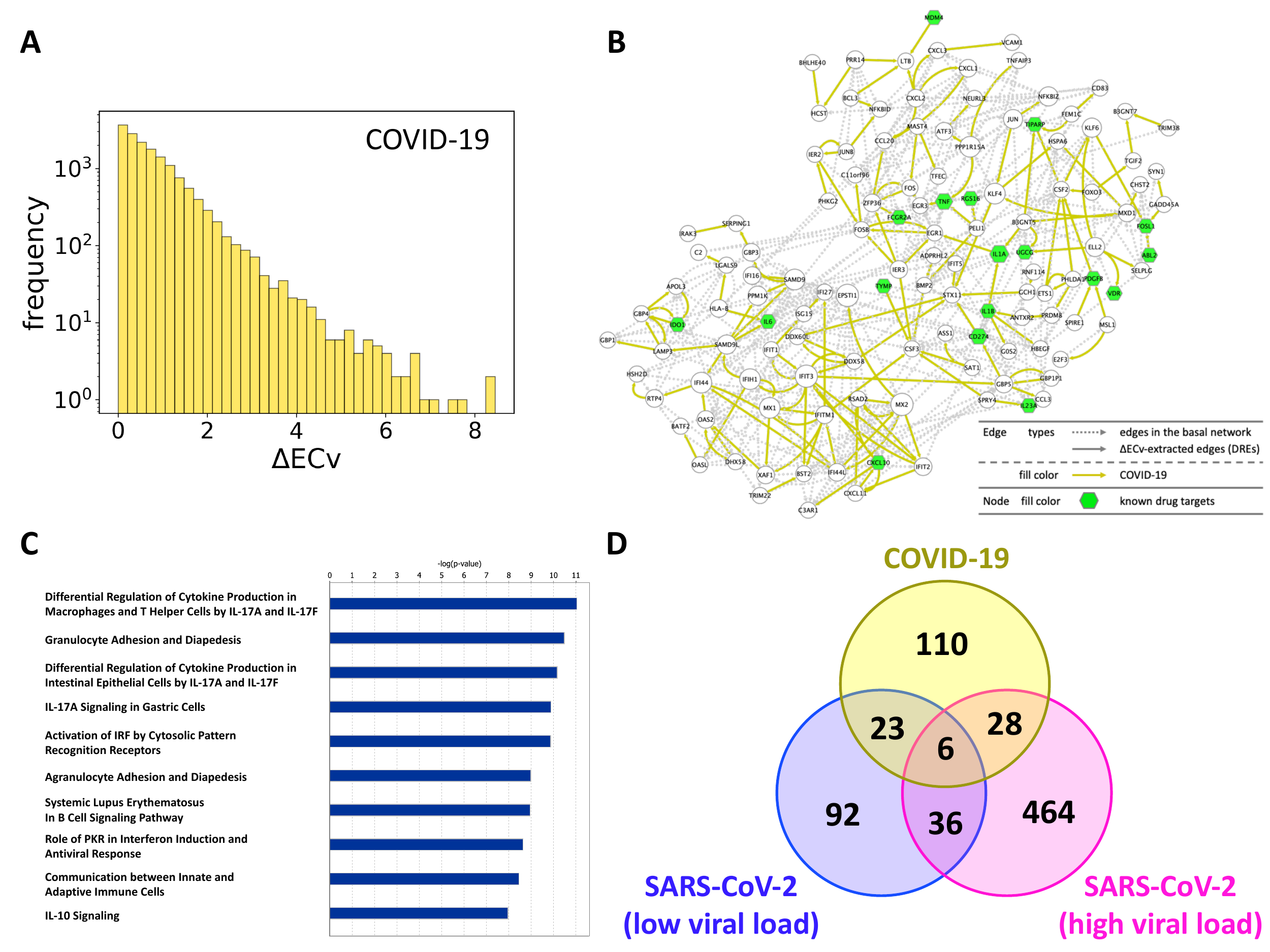}
\end{center}
\caption{The COVID-19-perturbated network analysis.
(\textbf{A}) The histograms of $\Delta$ECv for the biopsy dataset. The X-axis corresponds to the threshold for the $\Delta$ECv. The Y-axis stands for the number of edges with log scale. (\textbf{B}) The COVID-19-perturbated network is shown. The network is composed of 127 nodes and 412 edges (including 245 basal edges). The colored solid edges represent DREs perturbated by COVID-19 (yellow). The dot edges represent the basal edges (grey). The nodes (green) represent the known drug target genes (Supplementary Table S1). The node size represents the extent of outdegree. (\textbf{C}) The top 10 terms of canonical pathway analysis for the genes in the COVID-19-perturbated network. (\textbf{D}) The Venn diagram shows the numbers of the $\Delta$ECv-extracted DREs ($\Delta$ECv threshold 2.3) induced by COVID-19 perturbation for biopsy dataset (yellow) overlapped with the two DREs through the Venn diagram analysis in Fig. 2B.}
\label{fig5}
\end{figure*}

Finally, we aimed to establish COVID-19 individual networks with a human biopsy dataset (healthy: two samples; COVID-19 positive: two samples) on the basis of the estimated basal network model. We expect that the {\it in vivo} biopsy dataset potentially provides a more clinically relevant perspective compared with the {\it in vitro} experiments. Usually, network estimation is impossible with such a small number of samples owing to the difficulty in acquisition of the robust network structure, yet our approach using the basal network model was capable of generating a context-specific network, even with a few samples of a different dataset (\textbf{Fig.~\ref{fig1}}). By using the $B$-spline regression model of the Bayesian network acquired by the estimation of the basal network, we first computed the ECv for the preprocessed biopsy dataset, despite the absence of some genes compared with the dataset used for the basal network estimation. To obtain DREs, we calculated $\Delta$ECv between healthy (regarded as control) and COVID-19 samples following the Eq.~(\ref{eq:ecv}) where $S$ = healthy ($|S|=2$) and $T$ = COVID-19 ($|T|=2$) (see Methods). The $\Delta$ECvs were distributed over a broad range, and 4242 DREs were observed at the threshold of 1 for $\Delta$ECv (\textbf{Fig.~\ref{fig5}A}). To extract more reliable DREs induced by COVID-19, we set the threshold of 2.3 corresponding approximately to $\log_2$FC where FC=5, which resulted in 638 DREs. These DREs were mapped as networks and the biggest connected component (167 DREs) was depicted with inclusion of the basal edges, generating the COVID-19-perturbated network, which comprised 127 nodes and 412 edges (\textbf{Fig.~\ref{fig5}B}). This network is supposedly a representation of the distinctive cellular system in patients with COVID-19. The pathway analysis of genes contained in this network showed that they were involved in the immune and inflammatory response (\textbf{Fig.~\ref{fig5}C}), supporting the consistency of our established network with the biological observations in COVID-19.

To determine the signatures of the acquired DREs in the COVID-19-perturbated network, we measured the ECv similarity for a set of the 167 DREs across the other experimental samples. This result showed that the ECv profiles in COVID-19 were most similar to the sample of high SARS-CoV-2 viral load in A549 cells in the vitro experiments (\textbf{Supplementary Fig. S2A}), further supporting that the obtained DREs were associated with SARS-CoV-2 infection. We further explored the extent to which the COVID-19-related 167 DREs overlapped with the Venn diagram established in \textbf{Fig.~\ref{fig2}B}. We observed that a moderate number of the DREs were shared by the cell models of SARS-CoV-2 perturbation (\textbf{Fig.~\ref{fig5}D}), and then these overlapped edges were mapped onto the COVID-19-perturbated network (\textbf{Supplementary Fig. S2B}). Unlike the network observations in \textbf{Fig.~\ref{fig3}}, we found that both the ISG-related webs (module 1) and subsequent cytokine signaling (module 3) involved in inflammatory cascades were coincidently present in the COVID-19-perturbated network, indicating that these two modules were continued to be mutually activated in COVID-19.

\begin{figure*}[p]
\begin{center}
\includegraphics[width=.8\linewidth]{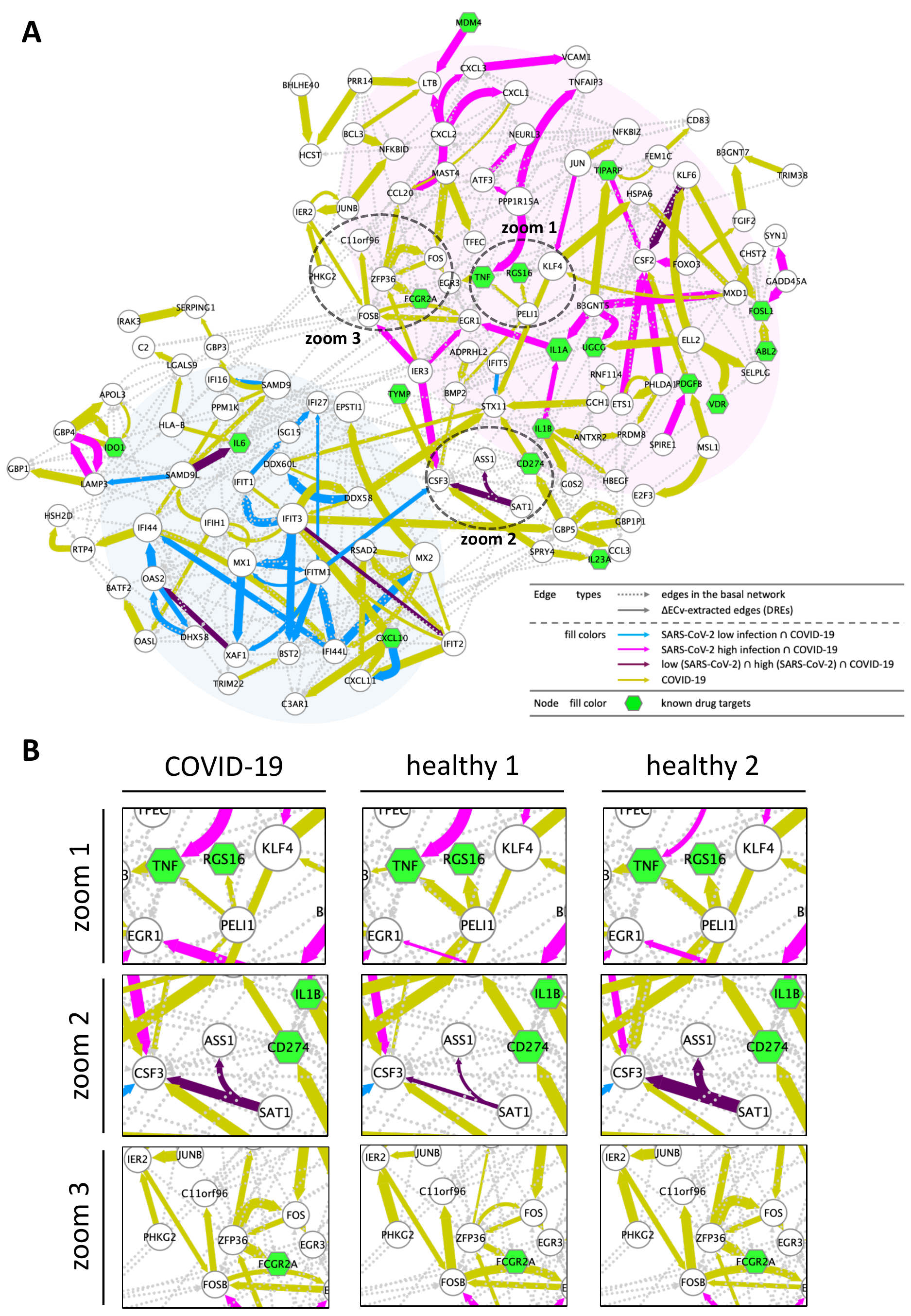}
\end{center}
\caption{Establishment of the COVID-19 patient-specific individual network.
(\textbf{A}) The COVID-19 patient-specific network with RCs represented by edge sizes. The network comprises 127 nodes and 412 edges (including 245 basal edges). The colored solid edges represent DREs; SARS-CoV-2 (high MOI: 2) $\cap$ COVID-19-perturbated (magenta), SARS-CoV-2 (lowMOI: 0.2) $\cap$ COVID-19-perturbated (blue), SARS-CoV-2 (high MOI: 2) $\cap$ SARS-CoV-2 (low MOI: 0.2) $\cap$ COVID-19-perturbated (purple), COVID-19-perturbated exclusive edges (yellow). The dot edges represent the basal edges (grey). The nodes (green) represents the known drug target genes (Supplementary Table S1). The node size stands for the extent of outdegree. (\textbf{B}) Zoomed regions indicated in Fig. 6A for three individuals (healthy 1, healthy 2, and the patient with COVID-19). RCs are represented as edge sizes to show individual differences.}
\label{fig6}
\end{figure*}

To uncover the differences in the local regulatory system, we next examined the profiles of the COVID-19-perturbated network at an individual level using the RC method (\textbf{Fig.~\ref{fig1}}). As the two COVID-19 samples were originally derived from a single patient who tested positive for COVID-19, we calculated RCs for three individuals (healthy 1, healthy 2, and COVID-19 patient). The depiction of the RC as the edge sizes eventually led to the establishment of the COVID-19 patient-specific network, which is likely to show how the cellular system changed in the patients with COVID-19 compared with the healthy controls (\textbf{Fig.~\ref{fig6}A}). The panel of three individual networks dramatically exhibited a great magnitude of differences, showing that the cellular regulatory systems were quite distinctive among individuals (\textbf{Supplementary Animation 2}). In comparison with the SARS-CoV-2-perturbated network established by the well-organized {\it in vitro} experiments using cell lines (\textbf{Supplementary Animation 1}), this broad range of RC fluctuation for each {\it in vitro} sample probably reflects further differences among individuals. The representative regions where local regulatory systems are different among individuals were illustrated in \textbf{Fig.~\ref{fig6}B}. In the zoom 1 region, PELI1 is a parent gene for both TNF and RGS16; these two signals were dominant in the healthy individuals, but not in the patient with COVID-19. In contrast, the zoom 2 and 3 regions showed that local signals were clearly different, not only between the healthy patients and the patient with COVID-19, but also even between two healthy individuals.




\section*{Discussion}

Here, we have presented the host cellular gene networks perturbated by SARS-CoV-2 both {\it in vitro} and {\it in vivo} by using our proposed framework for gene network analysis. As the networks we established to be associated with SARS-CoV-2 were generated through RNA-seq data, these networks explain how genes are systematically regulated at the transcriptome level. Although our approach depends on the initial network estimation with an experimental dataset and may therefore risk the inclusion of false relationships or the exclusion of true relationships, we have succeeded in capturing the biologically explainable immune response systems in human cells induced by SARS-CoV-2 at the level of signaling networks.

Sensing viruses cause an immune defense system in host cells, and this induces acute IFN signaling activation followed by expression of IFNs. These IFNs amplify JAK/STAT signaling to promote the expression of various ISGs and accelerate subsequent cytokine signaling \cite{Mesev2019-za}. As illustrated in \textbf{Fig.~\ref{fig3}}, the mutually interacting module of ISGs (module 1) followed by IFN and JAK/STAT signaling was shown to be an early response to SARS-CoV-2 infection. During the process of cells exposed to high SARS-CoV-2 viral loads, the signaling appears to move to the next stage, represented by inflammatory signaling, including the involvement of various cytokines (\textbf{Fig.~\ref{fig3}}). The recent reported drug, dexamethasone, could be effective for severe patients with COVID-19 by suppressing these orchestrated inflammatory signaling cascades \cite{RECOVERY_Collaborative_Group2020-bs}. In this network, IL6 was located as a hub gene to regulate downstream cascades, including chemokines and colony-stimulating factors, which are reported to be increased in patients with COVID-19 \cite{Huang2020-vp}. The web of chemokines, such as CXCL1, CXCL2, and CXCL3, may represent how SARS-CoV-2-infected cells present a signal to induce leukocyte chemotaxis and infiltration. The localization of ICAM1 in the vicinity of IL6 and chemokines is supportive of this, as ICAM1 is known to be a scaffold for the accumulation of leukocytes at inflammatory sites and its expression is regulated by cytokines, including IL6 \cite{Meager1999-nm,Velazquez-Salinas2019-pz}. This tendency was also observed in the network comparison analyses across four respiratory viruses, including SARS-CoV-2 (\textbf{Supplementary Fig. S1C}). These data showed that IL6 was not exclusive to SARS-CoV-2, but a universal factor in response to respiratory viral infection, except for influenza A virus. Given that several studies have reported that tocilizumab, an inhibitor of the IL6 receptor, is a potential drug able to suppress the cytokine storm observed in many critical patients with COVID-19 \cite{Fu2020-nx,Xu2020-dc}, the accumulated evidence strongly suggested that IL6 would be a central regulator of the inflammatory cascade, even from a network perspective. In addition, our network showed that CSF2 is regulated via various factors, including IL6, which strengthens previous reports suggesting that CSF2 may be a promising therapeutic target in combination with IL6 \cite{Zhou2020-to,Lang2020-cd}. Several recent studies have shown that ACE2 plays a key role in the process of SARS-CoV-2 infection. SARS-CoV-2 enters into host cells via ACE2 \cite{Hoffmann2020-da}, and ACE2 was found to be an ISG in human airway epithelial cells \cite{Ziegler2020-pb}. Considering that the SARS-CoV-2-perturbated network includes several ISGs (\textbf{Fig.~\ref{fig3}}), it can be reasoned that some clues regarding ACE2 may be present in this network. In this context, we found that ACE2 was closely located to this network and was downstream of TNFRSF9, ATF3, and ARRDC3 via ACHE (data not shown); these are potential candidates for further investigation of the relationship between ACE2 and ISGs. Thus, our networks provide promising information to elucidate SARS-CoV-2 profiles from a broad biological perspective. 

Our second noteworthy outcome in this study was that we succeeded in the characterization of sample-specific individual networks by introducing the new edge-quantitative technique of RC. In particular, although it is impossible to estimate a network with a small number of samples, such as the four biopsy samples used in our case, the basal network model that was already obtained through the analysis of the {\it in vitro} dataset with both RC and ECv methods has led to establishment of the COVID-19 patient-specific individual network. This process represents how we extrapolate between {\it in vitro} and {\it in vivo} experiments. Different samples each exhibit a unique regulatory profile, especially in actual individuals (such as those obtained from biopsy) rather than well-controlled {\it in vitro} samples (\textbf{Supplementary Animation 1 and 2}). These results probably reflect a more realistic clinical situation and increase the importance of making the most of effective utilization of a biopsy dataset. In the current outbreak of COVID-19, we need to look into both biological and clinical aspects for exploring COVID-19 therapy. The individual networks regarding COVID-19 shows the extent to which individuals possess their own network, which ultimately links to the necessity of personalized treatment. Therefore, our efforts are a potential contribution to the emerging field of personalized medicine. The biopsy dataset that we used was not sufficient to allow interpretation of the comprehensive information through individual networks in patients with COVID-19, as it contained fewer COVID-19 samples. More clinical samples from patients with COVID-19 can lead to the determination of key regulatory systems at a clinical level. We hope that our panel of network analyses will be of help to the SARS-CoV-2 research field and to establishment further treatments for COVID-19.

\section*{Methods}

\subsection*{Global gene network estimation and core network extraction}

In general, methods for gene network analysis are intended for the extraction of gene-to-gene regulatory relationships universally underlying given transcriptome datasets. Unlike commonly existing gene networks, we recently developed a method to extract sample-specific gene networks. Our method first estimates a global gene network, called the {\it basal network}, that included all the genes in a dataset using a Bayesian network with $B$-spline nonparametric regression \cite{Imoto2002-wm,Tamada2011-ub}. Bayesian network estimation is capable of capturing global cause-and-effect relationships among gene expression, rather than extracting locally co-regulated genes, such as co-expression correlation networks. This is realized by finding the conditional independencies among variables. In gene network analysis using a Bayesian network, gene expression is regarded as an observed sample from the random variables that correspond to genes or transcriptomes in a cell. 

Let $X_1,\ldots,X_p$ be $p$ variables of genes. In a Bayesian network, we consider the joint density of $p$ variables and assume that it is decomposed as the product of local conditional densities, such that
\begin{equation*}
f(X_1,\ldots,X_p; \theta_G)
        	= \prod_{j=1}^{p} f(X_j|X_{j_{1}},\ldots,X_{j_{q_j}}; \theta_j),
\end{equation*}
where $j_1,\dots,j_{q_j}$ are indices of $q_j$ dependent variables of the $j$-th variable. This decomposition can be represented as a directed acyclic graph (network) and variables $X_{j_1}, \ldots, X_{j_{q_j}}$ are connected as parents or inputs of the $j$-th variable in the network.

The $B$-spline nonparametric regression version of the Bayesian network models gene-to-gene expression relationships as mathematical equations using $B$-spline curves, such that
\begin{equation}
x_j = m^{(j)}_{1}(x_{j_1}) + \cdots + m^{(j)}_{q_j}(x_{j_{q_j}}) + \varepsilon_j,
\label{eq:regression}
\end{equation}
where $x_j$ represents an expression value of the $j$-th gene, $\varepsilon_j$ is the error term normally distributed with mean $0$ and variance $\sigma_j$, and $m^{(j)}_{k}(x) = \sum_{m=1}^{M} \gamma_{jkm} b_{jkm}(x)$ is a regression function using $M$ $B$-spline basis functions $b_{jkm}(\cdot)$, and their coefficients $\gamma_{jkm}$.

The structure search of the Bayesian network corresponds to finding the decomposition of the joint density. This is implemented by the maximization of the posterior probability, such that
\begin{equation*}
p(G|X) \propto \pi(G) \int
	\prod_{i=1}^{n} f(x_{i1},\cdots,x_{ip}; \theta_G) \pi(\theta_G|\lambda) d \theta_G,
\end{equation*}
where $X$ is an $n$-by-$p$ input matrix whose element $x_{ij}$ corresponds to an expression value of the $i$-th sample for the $j$-th gene, $G$ represents the network structure, $\pi(G)$ is the prior probability of $G$, $\theta_G$ is the parameter vector of the local conditional densities, $\pi(\theta_G|\lambda)$ is the prior distribution of $\theta_G$, and $\lambda$ is the hyperparameter vector. The difficulty in gene network estimation by the Bayesian network is the step of structure learning for large networks, because this is known to be an NP-hard problem; namely, an exponential increase in search space to the number of variables. We have used the neighbor node sampling and repeat algorithm that realizes the estimation of the large Bayesian network structure \cite{Tamada2011-ub}. It repeats the subnetwork estimation many times in parallel for the sampled variable sets by random walking, and thus it can estimate the large network within a realistic time.
 
After the basal network estimation, we then quantified every single edge with respect to a certain sample in terms of the system-level usage of the edge with the estimated mathematical model. Tanaka et al. \cite{Tanaka2020-xr} defined an {\it edge contribution value} (ECv) of edge $j_k \rightarrow j$ as \begin{equation*}
ECv_{(u)}(j_k \rightarrow j) = m^{(j)}_{k}(x^{(u)}_{j_k})
\end{equation*}
where $x^{(u)}_{j_k}$ represented the expression value of the $j_k$-th gene at a certain sample denoted by $u$, and $m^{(j)}_{k}(\cdot)$ was a regression function defined in Eq.~(\ref{eq:regression}). Note that sample $u$ did not necessarily have to be a single sample for use in the network estimation. They proved that ECv can be used for the quantification of edge $j_k \rightarrow j$ with respect to a given sample. To extract sample-specific networks, they considered the differences of ECvs between two different conditions of samples, similar to extracting differentially expressed genes by comparing control and perturbated expressions. They defined $\Delta$ECv as

\begin{equation}
\begin{split}
&\Delta ECv_{(S,T)}(j_{k}\rightarrow j) = \\
&\left| \frac{1}{|S|} \sum_{s \in S} ECv_{(s)}(j_{k}\rightarrow j) - \frac{1}{|T|} \sum_{t \in T} ECv_{(t)}(j_{k}\rightarrow j) \right|, \label{eq:ecv}
\end{split}
\end{equation}

where $S$ and $T$ are sets of samples observed in the particular conditions, respectively, in which $|S| \ge 1$ and $|T| \ge 1$. Note that in the case of $|S|=|T|=1$, this allows consideration of the differences between just two samples, e.g. control and perturbated samples. In general, we assume multiple replicated samples or a set of individual samples for both $S$ and $T$. By extracting edges and their connected nodes with $\Delta$ECvs greater or equal to a certain threshold, we can define the sample- or condition-specific core network from the basal network. In this study, $S$ and $T$ were sets of infected and control (mock) replicated samples, respectively. As the target dataset includes control samples for a particular series of experiments, we can extract certain core networks from them by calculating $\Delta$ECv for the series of experiments using their corresponding control samples. For example, we extracted a SARS-CoV-2-perturbated core network by calculating $\Delta$ECvs for SARS-CoV-2-infected and their corresponding mock-triplicate samples. As performed in the previous study, we generally employed $\Delta\textrm{ECv} \ge 1.0$ for the threshold of the core network extraction. This approximately corresponds to 2-fold changes in differentially expressed genes for extracted genes. Thus, we considered the extracted networks, including edges and nodes, which showed the significant activation of regulatory systems by the infection.

\subsection*{Proposed relative contribution of edges for characterizing individual networks}

This ECv development allowed a new solution for gene network analyses. In the previous study \cite{Tanaka2020-xr}, we succeeded in characterizing network profiles by calculating ECvs for edges in a $\Delta$ECv-extracted core network with respect to many samples from patients with cancer. The conventional clustering onto these calculated ECvs led to the identification of prognosis-related subgroups. Thus, we demonstrated that the differences and similarities in edge profiles of the network could be captured as patterns of ECvs. Despite the high availability of ECvs, it is still impossible to directly compare ECvs between individual samples, because ECvs have different sizes depending on the estimated pairwise edge and the sample. The normalization of ECvs across samples is inappropriate for our purpose owing to the mutual dependency of the individual network on each sample. Thus, it has not been possible to highlight the differences in regulatory systems at an individual level.

For these reasons, ECvs are not appropriate to the analysis of individual networks. To overcome the drawbacks, we have proposed a novel method, {\it relative contribution} (RC), to quantify edges with respect to individual samples using the estimated gene network model. We hypothesized that the differences of individual samples in terms of the cellular system can be observed as the differences in ratios of the contributions of edges connecting to a certain node in the network. Edges with different samples need to be described as differently weighted edges according to ratios of effects between parents that regulate or are connected to a certain gene. In addition, the quantification of a network with a single sample needs to be independent from other samples and their distributions. To realize this, we define the relative contribution of an edge with respect to a sample as
\begin{equation*}
RC_{(u)}(j_k \rightarrow j)
  = \frac{|ECv_{(u)}(j_k \rightarrow j)|}
  	{\max_{1 \le k' \le q_j} |ECv_{(u)}(j_{k'} \rightarrow j)|},
\end{equation*}
where $u$ represents a certain sample ($0 < RC \leq 1$). That is, an RC of the edge is a relative strength of the contribution of the edge to the maximum strength among the parents connecting to the same child node. The reason why an RC is not divided by the sum of the ECvs is that the range of RCs does not shrink depending on the number of parents of the child node. A drawback of RCs is that, if the ratio of ECvs of the parents is not changed, the changes of parent values do not affect the RCs. However, RCs of their downstreams will be affected by such changes. Therefore, this drawback is not problematic in terms of the specification of differences in individual networks. Note that, similar to ECvs, sample $u$ does not necessarily need to be a single sample used for the network estimation. As illustrated in Results, we have shown that RCs can be used to analyze individual networks, even if we have a single sample, or only a few samples, of gene expression data, as long as a basal network can be estimated from other datasets. RC therefore offers a significant enhancement to our framework for gene network analysis. Our data have demonstrated that the framework, through an integration of the three key pieces – Bayesian network estimation, ECv, and RC – provides a powerful data-driven solution to seek biological phenomena through cellular systems ranging from a global level to an individual level. Our proposed framework is mathematically
 illustrated in \textbf{Supplementary Fig. S3}.

\subsection*{Dataset}
The transcriptome dataset GSE147507 was downloaded from NCBI Gene Expression Omnibus
\cite{Blanco-Melo2020-cv}. The samples were infected with respiratory viruses, including SARS-CoV-2, and biological replicates were performed. We first selected samples exclusive for human RNA-seq with 78 samples. Among the samples, four samples of the {\it in vivo} experiment (biopsy) data were pre-eliminated. The $\log_2$-transformed dataset was filtered to remove genes with a mean percentile lower than 30$\%$, resulting in 74 samples and 15,258 genes. This preprocessed dataset of the 74 $\times$ 15258 matrix was used as input for the basal network estimation. The biopsy dataset eliminated above prior to global network estimation consisted of four samples (two healthy samples and two COVID-19-positive samples). The RPM (reads per million) normalized biopsy dataset was $\log_2$-transformed and genes with at least one zero value were removed to obtain more reliable data. The two technical replicate samples for COVID-19 were averaged for RC calculation. Following this preprocessing, the input dataset for RC calculation finally comprised a 3 $\times$ 4516 matrix. The RNA-seq samples used for $\Delta$ECv calculations in this study were: SARS-CoV-2 in A549 cells (MOI of 0.2/2 for 24 hr, n=3) and the corresponding mock (n=3); SARS-CoV-2 in normal human bronchial epithelial (NHBE) cells (MOI of 2 for 24 hr, n=3) and the corresponding mock (n=3); SARS-CoV-2 in Calu-3 cells (MOI of 2 for 24 hr, n=3) and the corresponding mock (n=3); human respiratory syncytial virus (RSV) in A549 cells (MOI of 2 for 24 hr, n=3) and the corresponding mock (n=3); human parainfluenza virus 3 (HPIV3) in A549 cells (MOI of 2 for 24 hr, n=3) and the corresponding mock (n=3); influenza A virus (IAV) in A549 cells (MOI of 5 for 9 hr, n=2) and the corresponding mock (n=2); COVID-19 (n=2) and healthy (n=2).

\subsection*{Pathway analysis}
The canonical pathway analysis was performed through the use of Ingenuity Pathway Analysis software \cite{Kramer2014-mi}.

\subsection*{Network analysis and visualization}
The network visualization and the network analysis were performed using Cytoscape (version 3.7.2 and 3.8.0) \cite{Shannon2003-va}. The genes for known drug targets were acquired from IPA knowledge database \cite{Kramer2014-mi} and the representative drugs were listed in \textbf{Supplementary Table S1}.

\subsection*{Computer environments}
All the computation for the network estimation and the ECv calculations in this study were performed by the SHIROKANE supercomputer system (Shirokane5) at Human Genome Center, the Institute of Medical Science, the University of Tokyo, where the computation nodes were equipped with dual Intel Xeon Gold 6154 3.0GHz CPUs and 192GB memory per node.

\subsection*{Data availability}
All the network files generated in this study are provided on the supplementary data (\href{http://ytlab.jp/suppl/tanaka_arxiv2020/index.html}{http://ytlab.jp/suppl/tanaka\_arxiv2020/index.html}). The program for network estimation is freely available for SHIROKANE users. The ECv/RC calculation program is available for non-commercial academic users upon request.

\begin{acknowledgements}
We thank M. Ikeguchi for technical support. The super-computing resource was provided by Human Genome Center, the Institute of Medical Science, the University of Tokyo. This work was supported by RIKEN Junior Research Associate Program. This study was performed using funds provided by PRISM (Public/Private R$\&$D Investment Strategic Expansion PrograM) in Japan. This paper format was generated through self-modification of the original template designed by Ricardo Henriques.
\end{acknowledgements}

\begin{contributions}
Y.T. (Yoshihisa Tanaka), Y.T.* (Yoshinori Tamada), and Y.O. conceived the experiments, Y.T., K.H., and M.A.N. conducted the experiments, K.H. visualized the networks, Y.T., Y.T.*, and Y.O. analyzed the results, Y.T. and Y.T.* wrote the manuscript, M.A.N., F.Y., and Y.O. reviewed and edited the manuscript.
\end{contributions}

\begin{interests}
Y. Tamada and Y. Okuno have a patent application on the individual profiling and the method for extraction of a sample-specific network used in this study through the technology licensing organization in Kyoto University. Other authors declare no conflict of interest.
\end{interests}

\section*{Bibliography}
\bibliography{covid19.bib}

\begin{thebibliography}{39}
\providecommand{\natexlab}[1]{#1}
\providecommand{\url}[1]{\texttt{#1}}
\expandafter\ifx\csname urlstyle\endcsname\relax
  \providecommand{\doi}[1]{doi: #1}\else
  \providecommand{\doi}{doi: \begingroup \urlstyle{rm}\Url}\fi

\bibitem[Zhu et~al.(2020)Zhu, Zhang, Wang, Li, Yang, Song, Zhao, Huang, Shi,
  Lu, Niu, Zhan, Ma, Wang, Xu, Wu, Gao, Tan, and {China Novel Coronavirus
  Investigating and Research Team}]{Zhu2020-wm}
Na~Zhu, Dingyu Zhang, Wenling Wang, Xingwang Li, Bo~Yang, Jingdong Song, Xiang
  Zhao, Baoying Huang, Weifeng Shi, Roujian Lu, Peihua Niu, Faxian Zhan, Xuejun
  Ma, Dayan Wang, Wenbo Xu, Guizhen Wu, George~F Gao, Wenjie Tan, and {China
  Novel Coronavirus Investigating and Research Team}.
\newblock A novel coronavirus from patients with pneumonia in china, 2019.
\newblock \emph{N. Engl. J. Med.}, 382\penalty0 (8):\penalty0 727--733,
  February 2020.
\newblock ISSN 0028-4793, 1533-4406.
\newblock \doi{10.1056/NEJMoa2001017}.

\bibitem[Wu et~al.(2020)Wu, Zhao, Yu, Chen, Wang, Song, Hu, Tao, Tian, Pei,
  Yuan, Zhang, Dai, Liu, Wang, Zheng, Xu, Holmes, and Zhang]{Wu2020-tu}
Fan Wu, Su~Zhao, Bin Yu, Yan-Mei Chen, Wen Wang, Zhi-Gang Song, Yi~Hu, Zhao-Wu
  Tao, Jun-Hua Tian, Yuan-Yuan Pei, Ming-Li Yuan, Yu-Ling Zhang, Fa-Hui Dai,
  Yi~Liu, Qi-Min Wang, Jiao-Jiao Zheng, Lin Xu, Edward~C Holmes, and Yong-Zhen
  Zhang.
\newblock A new coronavirus associated with human respiratory disease in china.
\newblock \emph{Nature}, 579\penalty0 (7798):\penalty0 265--269, March 2020.
\newblock ISSN 0028-0836, 1476-4687.
\newblock \doi{10.1038/s41586-020-2008-3}.

\bibitem[Dong et~al.(2020)Dong, Du, and Gardner]{Dong2020-xn}
Ensheng Dong, Hongru Du, and Lauren Gardner.
\newblock An interactive web-based dashboard to track {COVID-19} in real time.
\newblock \emph{Lancet Infect. Dis.}, 20\penalty0 (5):\penalty0 533--534, May
  2020.
\newblock ISSN 1473-3099, 1474-4457.
\newblock \doi{10.1016/S1473-3099(20)30120-1}.

\bibitem[Hellewell et~al.(2020)Hellewell, Abbott, Gimma, Bosse, Jarvis,
  Russell, Munday, Kucharski, Edmunds, {Centre for the Mathematical Modelling
  of Infectious Diseases COVID-19 Working Group}, Funk, and
  Eggo]{Hellewell2020-np}
Joel Hellewell, Sam Abbott, Amy Gimma, Nikos~I Bosse, Christopher~I Jarvis,
  Timothy~W Russell, James~D Munday, Adam~J Kucharski, W~John Edmunds, {Centre
  for the Mathematical Modelling of Infectious Diseases COVID-19 Working
  Group}, Sebastian Funk, and Rosalind~M Eggo.
\newblock Feasibility of controlling {COVID-19} outbreaks by isolation of cases
  and contacts.
\newblock \emph{Lancet Glob Health}, 8\penalty0 (4):\penalty0 e488--e496, April
  2020.
\newblock ISSN 2214-109X.
\newblock \doi{10.1016/S2214-109X(20)30074-7}.

\bibitem[Gordon et~al.(2020)Gordon, Jang, Bouhaddou, Xu, Obernier, White,
  O'Meara, Rezelj, Guo, Swaney, Tummino, H{\"u}ttenhain, Kaake, Richards,
  Tutuncuoglu, Foussard, Batra, Haas, Modak, Kim, Haas, Polacco, Braberg,
  Fabius, Eckhardt, Soucheray, Bennett, Cakir, McGregor, Li, Meyer, Roesch,
  Vallet, Mac~Kain, Miorin, Moreno, Naing, Zhou, Peng, Shi, Zhang, Shen, Kirby,
  Melnyk, Chorba, Lou, Dai, Barrio-Hernandez, Memon, Hernandez-Armenta, Lyu,
  Mathy, Perica, Pilla, Ganesan, Saltzberg, Rakesh, Liu, Rosenthal, Calviello,
  Venkataramanan, Liboy-Lugo, Lin, Huang, Liu, Wankowicz, Bohn, Safari, Ugur,
  Koh, Savar, Tran, Shengjuler, Fletcher, O'Neal, Cai, Chang, Broadhurst,
  Klippsten, Sharp, Wenzell, Kuzuoglu-Ozturk, Wang, Trenker, Young, Cavero,
  Hiatt, Roth, Rathore, Subramanian, Noack, Hubert, Stroud, Frankel, Rosenberg,
  Verba, Agard, Ott, Emerman, Jura, von Zastrow, Verdin, Ashworth, Schwartz,
  d'Enfert, Mukherjee, Jacobson, Malik, Fujimori, Ideker, Craik, Floor, Fraser,
  Gross, Sali, Roth, Ruggero, Taunton, Kortemme, Beltrao, Vignuzzi,
  Garc{\'\i}a-Sastre, Shokat, Shoichet, and Krogan]{Gordon2020-sf}
David~E Gordon, Gwendolyn~M Jang, Mehdi Bouhaddou, Jiewei Xu, Kirsten Obernier,
  Kris~M White, Matthew~J O'Meara, Veronica~V Rezelj, Jeffrey~Z Guo, Danielle~L
  Swaney, Tia~A Tummino, Ruth H{\"u}ttenhain, Robyn~M Kaake, Alicia~L Richards,
  Beril Tutuncuoglu, Helene Foussard, Jyoti Batra, Kelsey Haas, Maya Modak,
  Minkyu Kim, Paige Haas, Benjamin~J Polacco, Hannes Braberg, Jacqueline~M
  Fabius, Manon Eckhardt, Margaret Soucheray, Melanie~J Bennett, Merve Cakir,
  Michael~J McGregor, Qiongyu Li, Bjoern Meyer, Ferdinand Roesch, Thomas
  Vallet, Alice Mac~Kain, Lisa Miorin, Elena Moreno, Zun Zar~Chi Naing, Yuan
  Zhou, Shiming Peng, Ying Shi, Ziyang Zhang, Wenqi Shen, Ilsa~T Kirby, James~E
  Melnyk, John~S Chorba, Kevin Lou, Shizhong~A Dai, Inigo Barrio-Hernandez,
  Danish Memon, Claudia Hernandez-Armenta, Jiankun Lyu, Christopher J~P Mathy,
  Tina Perica, Kala~Bharath Pilla, Sai~J Ganesan, Daniel~J Saltzberg,
  Ramachandran Rakesh, Xi~Liu, Sara~B Rosenthal, Lorenzo Calviello, Srivats
  Venkataramanan, Jose Liboy-Lugo, Yizhu Lin, Xi-Ping Huang, Yongfeng Liu,
  Stephanie~A Wankowicz, Markus Bohn, Maliheh Safari, Fatima~S Ugur, Cassandra
  Koh, Nastaran~Sadat Savar, Quang~Dinh Tran, Djoshkun Shengjuler, Sabrina~J
  Fletcher, Michael~C O'Neal, Yiming Cai, Jason C~J Chang, David~J Broadhurst,
  Saker Klippsten, Phillip~P Sharp, Nicole~A Wenzell, Duygu Kuzuoglu-Ozturk,
  Hao-Yuan Wang, Raphael Trenker, Janet~M Young, Devin~A Cavero, Joseph Hiatt,
  Theodore~L Roth, Ujjwal Rathore, Advait Subramanian, Julia Noack, Mathieu
  Hubert, Robert~M Stroud, Alan~D Frankel, Oren~S Rosenberg, Kliment~A Verba,
  David~A Agard, Melanie Ott, Michael Emerman, Natalia Jura, Mark von Zastrow,
  Eric Verdin, Alan Ashworth, Olivier Schwartz, Christophe d'Enfert, Shaeri
  Mukherjee, Matt Jacobson, Harmit~S Malik, Danica~G Fujimori, Trey Ideker,
  Charles~S Craik, Stephen~N Floor, James~S Fraser, John~D Gross, Andrej Sali,
  Bryan~L Roth, Davide Ruggero, Jack Taunton, Tanja Kortemme, Pedro Beltrao,
  Marco Vignuzzi, Adolfo Garc{\'\i}a-Sastre, Kevan~M Shokat, Brian~K Shoichet,
  and Nevan~J Krogan.
\newblock A {SARS-CoV-2} protein interaction map reveals targets for drug
  repurposing.
\newblock \emph{Nature}, 583\penalty0 (7816):\penalty0 459--468, July 2020.
\newblock ISSN 0028-0836, 1476-4687.
\newblock \doi{10.1038/s41586-020-2286-9}.

\bibitem[Bojkova et~al.(2020)Bojkova, Klann, Koch, Widera, Krause, Ciesek,
  Cinatl, and M{\"u}nch]{Bojkova2020-hu}
Denisa Bojkova, Kevin Klann, Benjamin Koch, Marek Widera, David Krause, Sandra
  Ciesek, Jindrich Cinatl, and Christian M{\"u}nch.
\newblock Proteomics of {SARS-CoV-2-infected} host cells reveals therapy
  targets.
\newblock \emph{Nature}, 583\penalty0 (7816):\penalty0 469--472, July 2020.
\newblock ISSN 0028-0836, 1476-4687.
\newblock \doi{10.1038/s41586-020-2332-7}.

\bibitem[Blanco-Melo et~al.(2020)Blanco-Melo, Nilsson-Payant, Liu, Uhl,
  Hoagland, M{\o}ller, Jordan, Oishi, Panis, Sachs, Wang, Schwartz, Lim,
  Albrecht, and tenOever]{Blanco-Melo2020-cv}
Daniel Blanco-Melo, Benjamin~E Nilsson-Payant, Wen-Chun Liu, Skyler Uhl, Daisy
  Hoagland, Rasmus M{\o}ller, Tristan~X Jordan, Kohei Oishi, Maryline Panis,
  David Sachs, Taia~T Wang, Robert~E Schwartz, Jean~K Lim, Randy~A Albrecht,
  and Benjamin~R tenOever.
\newblock Imbalanced host response to {SARS-CoV-2} drives development of
  {COVID-19}.
\newblock \emph{Cell}, 181\penalty0 (5):\penalty0 1036--1045.e9, May 2020.
\newblock ISSN 0092-8674, 1097-4172.
\newblock \doi{10.1016/j.cell.2020.04.026}.

\bibitem[Yan et~al.(2018)Yan, Risacher, Shen, and Saykin]{Yan2018-hu}
Jingwen Yan, Shannon~L Risacher, Li~Shen, and Andrew~J Saykin.
\newblock Network approaches to systems biology analysis of complex disease:
  integrative methods for multi-omics data.
\newblock \emph{Brief. Bioinform.}, 19\penalty0 (6):\penalty0 1370--1381,
  November 2018.
\newblock ISSN 1467-5463, 1477-4054.
\newblock \doi{10.1093/bib/bbx066}.

\bibitem[Recanatini and Cabrelle(2020)]{Recanatini2020-jo}
Maurizio Recanatini and Chiara Cabrelle.
\newblock Drug research meets network science: Where are we?
\newblock \emph{J. Med. Chem.}, May 2020.
\newblock ISSN 0022-2623, 1520-4804.
\newblock \doi{10.1021/acs.jmedchem.9b01989}.

\bibitem[Guzzi et~al.(2020)Guzzi, Mercatelli, Ceraolo, and
  Giorgi]{Guzzi2020-rr}
Pietro~H Guzzi, Daniele Mercatelli, Carmine Ceraolo, and Federico~M Giorgi.
\newblock Master regulator analysis of the {SARS-CoV-2/Human} interactome.
\newblock \emph{J. Clin. Med. Res.}, 9\penalty0 (4), April 2020.
\newblock ISSN 1918-3003, 2077-0383.
\newblock \doi{10.3390/jcm9040982}.

\bibitem[Zhou et~al.(2020{\natexlab{a}})Zhou, Hou, Shen, Huang, Martin, and
  Cheng]{Zhou2020-uf}
Yadi Zhou, Yuan Hou, Jiayu Shen, Yin Huang, William Martin, and Feixiong Cheng.
\newblock Network-based drug repurposing for novel coronavirus
  {2019-nCoV/SARS-CoV-2}.
\newblock \emph{Cell Discov}, 6:\penalty0 14, March 2020{\natexlab{a}}.
\newblock ISSN 2056-5968.
\newblock \doi{10.1038/s41421-020-0153-3}.

\bibitem[Gysi et~al.(2020)Gysi, Do~Valle, Zitnik, Ameli, Gan, Varol, Sanchez,
  Baron, Ghiassian, Loscalzo, and Barab{\'a}si]{Gysi2020-pd}
Deisy~Morselli Gysi, {\'I}talo Do~Valle, Marinka Zitnik, Asher Ameli, Xiao Gan,
  Onur Varol, Helia Sanchez, Rebecca~Marlene Baron, Dina Ghiassian, Joseph
  Loscalzo, and Albert-L{\'a}szl{\'o} Barab{\'a}si.
\newblock Network medicine framework for identifying drug repurposing
  opportunities for {COVID-19}.
\newblock \emph{ArXiv}, April 2020.
\newblock ISSN 2331-8422.

\bibitem[Fagone et~al.(2020)Fagone, Ciurleo, Lombardo, Iacobello, Palermo,
  Shoenfeld, Bendtzen, Bramanti, and Nicoletti]{Fagone2020-nz}
Paolo Fagone, Rosella Ciurleo, Salvo~Danilo Lombardo, Carmelo Iacobello,
  Concetta~Ilenia Palermo, Yehuda Shoenfeld, Klaus Bendtzen, Placido Bramanti,
  and Ferdinando Nicoletti.
\newblock Transcriptional landscape of {SARS-CoV-2} infection dismantles
  pathogenic pathways activated by the virus, proposes unique sex-specific
  differences and predicts tailored therapeutic strategies.
\newblock \emph{Autoimmun. Rev.}, page 102571, May 2020.
\newblock ISSN 1568-9972, 1873-0183.
\newblock \doi{10.1016/j.autrev.2020.102571}.

\bibitem[Tanaka et~al.(2020)Tanaka, Tamada, Ikeguchi, Yamashita, and
  Okuno]{Tanaka2020-xr}
Yoshihisa Tanaka, Yoshinori Tamada, Marie Ikeguchi, Fumiyoshi Yamashita, and
  Yasushi Okuno.
\newblock {System-Based} differential gene network analysis for characterizing
  a {Sample-Specific} subnetwork.
\newblock \emph{Biomolecules}, 10\penalty0 (2), February 2020.
\newblock ISSN 2218-273X.
\newblock \doi{10.3390/biom10020306}.

\bibitem[Margolin et~al.(2006)Margolin, Nemenman, Basso, Wiggins, Stolovitzky,
  Dalla~Favera, and Califano]{Margolin2006-oc}
Adam~A Margolin, Ilya Nemenman, Katia Basso, Chris Wiggins, Gustavo
  Stolovitzky, Riccardo Dalla~Favera, and Andrea Califano.
\newblock {ARACNE}: an algorithm for the reconstruction of gene regulatory
  networks in a mammalian cellular context.
\newblock \emph{BMC Bioinformatics}, 7 Suppl 1:\penalty0 S7, March 2006.
\newblock ISSN 1471-2105.
\newblock \doi{10.1186/1471-2105-7-S1-S7}.

\bibitem[Araki et~al.(2009)Araki, Tamada, Imoto, Dunmore, Sanders, Humphrey,
  Nagasaki, Doi, Nakanishi, Yasuda, Tomiyasu, Tashiro, Print, Stephen
  Charnock-Jones, Kuhara, and Miyano]{Araki2009-lf}
Hiromitsu Araki, Yoshinori Tamada, Seiya Imoto, Ben Dunmore, Deborah Sanders,
  Sally Humphrey, Masao Nagasaki, Atsushi Doi, Yukiko Nakanishi, Kaori Yasuda,
  Yuki Tomiyasu, Kousuke Tashiro, Cristin Print, D~Stephen Charnock-Jones,
  Satoru Kuhara, and Satoru Miyano.
\newblock Analysis of {PPAR$\alpha$-dependent} and {PPAR$\alpha$-independent}
  transcript regulation following fenofibrate treatment of human endothelial
  cells.
\newblock \emph{Angiogenesis}, 12\penalty0 (3):\penalty0 221--229, September
  2009.
\newblock ISSN 0969-6970, 1573-7209.
\newblock \doi{10.1007/s10456-009-9142-8}.

\bibitem[Wang et~al.(2012)Wang, Hurley, Watkins, Araki, Tamada, Muthukaruppan,
  Ranjard, Derkac, Imoto, Miyano, Crampin, and Print]{Wang2012-qb}
Li~Wang, Daniel~G Hurley, Wendy Watkins, Hiromitsu Araki, Yoshinori Tamada,
  Anita Muthukaruppan, Louis Ranjard, Eliane Derkac, Seiya Imoto, Satoru
  Miyano, Edmund~J Crampin, and Cristin~G Print.
\newblock Cell cycle gene networks are associated with melanoma prognosis.
\newblock \emph{PLoS One}, 7\penalty0 (4):\penalty0 e34247, April 2012.
\newblock ISSN 1932-6203.
\newblock \doi{10.1371/journal.pone.0034247}.

\bibitem[Affara et~al.(2013)Affara, Sanders, Araki, Tamada, Dunmore, Humphreys,
  Imoto, Savoie, Miyano, Kuhara, Jeffries, Print, and
  Charnock-Jones]{Affara2013-vc}
Muna Affara, Debbie Sanders, Hiromitsu Araki, Yoshinori Tamada, Benjamin~J
  Dunmore, Sally Humphreys, Seiya Imoto, Christopher Savoie, Satoru Miyano,
  Satoru Kuhara, David Jeffries, Cristin Print, and D~Stephen Charnock-Jones.
\newblock Vasohibin-1 is identified as a master-regulator of endothelial cell
  apoptosis using gene network analysis.
\newblock \emph{BMC Genomics}, 14:\penalty0 23, January 2013.
\newblock ISSN 1471-2164.
\newblock \doi{10.1186/1471-2164-14-23}.

\bibitem[Singh et~al.(2018)Singh, Ramsey, Filtz, and Kioussi]{Singh2018-sz}
Arun~J Singh, Stephen~A Ramsey, Theresa~M Filtz, and Chrissa Kioussi.
\newblock Differential gene regulatory networks in development and disease.
\newblock \emph{Cell. Mol. Life Sci.}, 75\penalty0 (6):\penalty0 1013--1025,
  March 2018.
\newblock ISSN 1420-682X, 1420-9071.
\newblock \doi{10.1007/s00018-017-2679-6}.

\bibitem[Yan et~al.(2016)Yan, Wang, Sethi, Muir, Kitchen, Cheng, and
  Gerstein]{Yan2016-hu}
Koon-Kiu Yan, Daifeng Wang, Anurag Sethi, Paul Muir, Robert Kitchen, Chao
  Cheng, and Mark Gerstein.
\newblock {Cross-Disciplinary} network comparison: Matchmaking between
  hairballs.
\newblock \emph{Cell Syst}, 2\penalty0 (3):\penalty0 147--157, March 2016.
\newblock ISSN 2405-4712.
\newblock \doi{10.1016/j.cels.2016.02.014}.

\bibitem[Huang et~al.(2020)Huang, Wang, Li, Ren, Zhao, Hu, Zhang, Fan, Xu, Gu,
  Cheng, Yu, Xia, Wei, Wu, Xie, Yin, Li, Liu, Xiao, Gao, Guo, Xie, Wang, Jiang,
  Gao, Jin, Wang, and Cao]{Huang2020-vp}
Chaolin Huang, Yeming Wang, Xingwang Li, Lili Ren, Jianping Zhao, Yi~Hu,
  Li~Zhang, Guohui Fan, Jiuyang Xu, Xiaoying Gu, Zhenshun Cheng, Ting Yu, Jiaan
  Xia, Yuan Wei, Wenjuan Wu, Xuelei Xie, Wen Yin, Hui Li, Min Liu, Yan Xiao,
  Hong Gao, Li~Guo, Jungang Xie, Guangfa Wang, Rongmeng Jiang, Zhancheng Gao,
  Qi~Jin, Jianwei Wang, and Bin Cao.
\newblock Clinical features of patients infected with 2019 novel coronavirus in
  wuhan, china.
\newblock \emph{Lancet}, 395\penalty0 (10223):\penalty0 497--506, February
  2020.
\newblock ISSN 0140-6736, 1474-547X.
\newblock \doi{10.1016/S0140-6736(20)30183-5}.

\bibitem[Guan et~al.(2020)Guan, Ni, Hu, Liang, Ou, He, Liu, Shan, Lei, Hui, Du,
  Li, Zeng, Yuen, Chen, Tang, Wang, Chen, Xiang, Li, Wang, Liang, Peng, Wei,
  Liu, Hu, Peng, Wang, Liu, Chen, Li, Zheng, Qiu, Luo, Ye, Zhu, Zhong, and
  {China Medical Treatment Expert Group for Covid-19}]{Guan2020-ci}
Wei-Jie Guan, Zheng-Yi Ni, Yu~Hu, Wen-Hua Liang, Chun-Quan Ou, Jian-Xing He,
  Lei Liu, Hong Shan, Chun-Liang Lei, David S~C Hui, Bin Du, Lan-Juan Li, Guang
  Zeng, Kwok-Yung Yuen, Ru-Chong Chen, Chun-Li Tang, Tao Wang, Ping-Yan Chen,
  Jie Xiang, Shi-Yue Li, Jin-Lin Wang, Zi-Jing Liang, Yi-Xiang Peng, Li~Wei,
  Yong Liu, Ya-Hua Hu, Peng Peng, Jian-Ming Wang, Ji-Yang Liu, Zhong Chen, Gang
  Li, Zhi-Jian Zheng, Shao-Qin Qiu, Jie Luo, Chang-Jiang Ye, Shao-Yong Zhu,
  Nan-Shan Zhong, and {China Medical Treatment Expert Group for Covid-19}.
\newblock Clinical characteristics of coronavirus disease 2019 in china.
\newblock \emph{N. Engl. J. Med.}, 382\penalty0 (18):\penalty0 1708--1720,
  April 2020.
\newblock ISSN 0028-4793, 1533-4406.
\newblock \doi{10.1056/NEJMoa2002032}.

\bibitem[Mehta et~al.(2020)Mehta, McAuley, Brown, Sanchez, Tattersall, Manson,
  and {HLH Across Speciality Collaboration, UK}]{Mehta2020-fc}
Puja Mehta, Daniel~F McAuley, Michael Brown, Emilie Sanchez, Rachel~S
  Tattersall, Jessica~J Manson, and {HLH Across Speciality Collaboration, UK}.
\newblock {COVID-19}: consider cytokine storm syndromes and immunosuppression.
\newblock \emph{Lancet}, 395\penalty0 (10229):\penalty0 1033--1034, March 2020.
\newblock ISSN 0140-6736, 1474-547X.
\newblock \doi{10.1016/S0140-6736(20)30628-0}.

\bibitem[Tamada et~al.(2011)Tamada, Imoto, Araki, Nagasaki, Print,
  Charnock-Jones, and Miyano]{Tamada2011-ub}
Yoshinori Tamada, Seiya Imoto, Hiromitsu Araki, Masao Nagasaki, Cristin Print,
  D~Stephen Charnock-Jones, and Satoru Miyano.
\newblock Estimating genome-wide gene networks using nonparametric bayesian
  network models on massively parallel computers.
\newblock \emph{IEEE/ACM Trans. Comput. Biol. Bioinform.}, 8\penalty0
  (3):\penalty0 683--697, May 2011.
\newblock ISSN 1545-5963, 1557-9964.
\newblock \doi{10.1109/TCBB.2010.68}.

\bibitem[Barab{\'a}si and Oltvai(2004)]{Barabasi2004-zv}
Albert-L{\'a}szl{\'o} Barab{\'a}si and Zolt{\'a}n~N Oltvai.
\newblock Network biology: understanding the cell's functional organization.
\newblock \emph{Nat. Rev. Genet.}, 5\penalty0 (2):\penalty0 101--113, February
  2004.
\newblock ISSN 1471-0056.
\newblock \doi{10.1038/nrg1272}.

\bibitem[Mesev et~al.(2019)Mesev, LeDesma, and Ploss]{Mesev2019-za}
Emily~V Mesev, Robert~A LeDesma, and Alexander Ploss.
\newblock Decoding type {I} and {III} interferon signalling during viral
  infection.
\newblock \emph{Nat Microbiol}, 4\penalty0 (6):\penalty0 914--924, June 2019.
\newblock ISSN 2058-5276.
\newblock \doi{10.1038/s41564-019-0421-x}.

\bibitem[Qian et~al.(2016)Qian, Xu, Zhao, and Qi]{Qian2016-oi}
Xijing Qian, Chen Xu, Ping Zhao, and Zhongtian Qi.
\newblock Long non-coding {RNA} {GAS5} inhibited hepatitis {C} virus
  replication by binding viral {NS3} protein.
\newblock \emph{Virology}, 492:\penalty0 155--165, May 2016.
\newblock ISSN 0042-6822, 1096-0341.
\newblock \doi{10.1016/j.virol.2016.02.020}.

\bibitem[{RECOVERY Collaborative Group} et~al.(2020){RECOVERY Collaborative
  Group}, Horby, Lim, Emberson, Mafham, Bell, Linsell, Staplin, Brightling,
  Ustianowski, Elmahi, Prudon, Green, Felton, Chadwick, Rege, Fegan, Chappell,
  Faust, Jaki, Jeffery, Montgomery, Rowan, Juszczak, Baillie, Haynes, and
  Landray]{RECOVERY_Collaborative_Group2020-bs}
{RECOVERY Collaborative Group}, Peter Horby, Wei~Shen Lim, Jonathan~R Emberson,
  Marion Mafham, Jennifer~L Bell, Louise Linsell, Natalie Staplin, Christopher
  Brightling, Andrew Ustianowski, Einas Elmahi, Benjamin Prudon, Christopher
  Green, Timothy Felton, David Chadwick, Kanchan Rege, Christopher Fegan,
  Lucy~C Chappell, Saul~N Faust, Thomas Jaki, Katie Jeffery, Alan Montgomery,
  Kathryn Rowan, Edmund Juszczak, J~Kenneth Baillie, Richard Haynes, and
  Martin~J Landray.
\newblock Dexamethasone in hospitalized patients with covid-19 - preliminary
  report.
\newblock \emph{N. Engl. J. Med.}, July 2020.
\newblock ISSN 0028-4793, 1533-4406.
\newblock \doi{10.1056/NEJMoa2021436}.

\bibitem[Meager(1999)]{Meager1999-nm}
A~Meager.
\newblock Cytokine regulation of cellular adhesion molecule expression in
  inflammation.
\newblock \emph{Cytokine Growth Factor Rev.}, 10\penalty0 (1):\penalty0 27--39,
  March 1999.
\newblock ISSN 1359-6101.
\newblock \doi{10.1016/s1359-6101(98)00024-0}.

\bibitem[Velazquez-Salinas et~al.(2019)Velazquez-Salinas, Verdugo-Rodriguez,
  Rodriguez, and Borca]{Velazquez-Salinas2019-pz}
Lauro Velazquez-Salinas, Antonio Verdugo-Rodriguez, Luis~L Rodriguez, and
  Manuel~V Borca.
\newblock The role of interleukin 6 during viral infections.
\newblock \emph{Front. Microbiol.}, 10:\penalty0 1057, May 2019.
\newblock ISSN 1664-302X.
\newblock \doi{10.3389/fmicb.2019.01057}.

\bibitem[Fu et~al.(2020)Fu, Xu, and Wei]{Fu2020-nx}
Binqing Fu, Xiaoling Xu, and Haiming Wei.
\newblock Why tocilizumab could be an effective treatment for severe
  {COVID-19}?
\newblock \emph{J. Transl. Med.}, 18\penalty0 (1):\penalty0 164, April 2020.
\newblock ISSN 1479-5876.
\newblock \doi{10.1186/s12967-020-02339-3}.

\bibitem[Xu et~al.(2020)Xu, Han, Li, Sun, Wang, Fu, Zhou, Zheng, Yang, Li,
  Zhang, Pan, and Wei]{Xu2020-dc}
Xiaoling Xu, Mingfeng Han, Tiantian Li, Wei Sun, Dongsheng Wang, Binqing Fu,
  Yonggang Zhou, Xiaohu Zheng, Yun Yang, Xiuyong Li, Xiaohua Zhang, Aijun Pan,
  and Haiming Wei.
\newblock Effective treatment of severe {COVID-19} patients with tocilizumab.
\newblock \emph{Proc. Natl. Acad. Sci. U. S. A.}, 117\penalty0 (20):\penalty0
  10970--10975, May 2020.
\newblock ISSN 0027-8424, 1091-6490.
\newblock \doi{10.1073/pnas.2005615117}.

\bibitem[Zhou et~al.(2020{\natexlab{b}})Zhou, Fu, Zheng, Wang, Zhao, Qi, Sun,
  Tian, Xu, and Wei]{Zhou2020-to}
Yonggang Zhou, Binqing Fu, Xiaohu Zheng, Dongsheng Wang, Changcheng Zhao,
  Yingjie Qi, Rui Sun, Zhigang Tian, Xiaoling Xu, and Haiming Wei.
\newblock Pathogenic t-cells and inflammatory monocytes incite inflammatory
  storms in severe {COVID-19} patients.
\newblock \emph{Natl Sci Rev}, 7\penalty0 (6):\penalty0 998--1002, June
  2020{\natexlab{b}}.
\newblock ISSN 2095-5138.
\newblock \doi{10.1093/nsr/nwaa041}.

\bibitem[Lang et~al.(2020)Lang, Lee, Teijaro, Becher, and
  Hamilton]{Lang2020-cd}
Frederick~M Lang, Kevin M-C Lee, John~R Teijaro, Burkhard Becher, and John~A
  Hamilton.
\newblock {GM-CSF-based} treatments in {COVID-19}: reconciling opposing
  therapeutic approaches.
\newblock \emph{Nat. Rev. Immunol.}, June 2020.
\newblock ISSN 1474-1733, 1474-1741.
\newblock \doi{10.1038/s41577-020-0357-7}.

\bibitem[Hoffmann et~al.(2020)Hoffmann, Kleine-Weber, Schroeder, Kr{\"u}ger,
  Herrler, Erichsen, Schiergens, Herrler, Wu, Nitsche, M{\"u}ller, Drosten, and
  P{\"o}hlmann]{Hoffmann2020-da}
Markus Hoffmann, Hannah Kleine-Weber, Simon Schroeder, Nadine Kr{\"u}ger, Tanja
  Herrler, Sandra Erichsen, Tobias~S Schiergens, Georg Herrler, Nai-Huei Wu,
  Andreas Nitsche, Marcel~A M{\"u}ller, Christian Drosten, and Stefan
  P{\"o}hlmann.
\newblock {SARS-CoV-2} cell entry depends on {ACE2} and {TMPRSS2} and is
  blocked by a clinically proven protease inhibitor.
\newblock \emph{Cell}, 181\penalty0 (2):\penalty0 271--280.e8, April 2020.
\newblock ISSN 0092-8674, 1097-4172.
\newblock \doi{10.1016/j.cell.2020.02.052}.

\bibitem[Ziegler et~al.(2020)Ziegler, Allon, Nyquist, Mbano, Miao, Tzouanas,
  Cao, Yousif, Bals, Hauser, Feldman, Muus, Wadsworth, Kazer, Hughes, Doran,
  Gatter, Vukovic, Taliaferro, Mead, Guo, Wang, Gras, Plaisant, Ansari,
  Angelidis, Adler, Sucre, Taylor, Lin, Waghray, Mitsialis, Dwyer, Buchheit,
  Boyce, Barrett, Laidlaw, Carroll, Colonna, Tkachev, Peterson, Yu, Zheng,
  Gideon, Winchell, Lin, Bingle, Snapper, Kropski, Theis, Schiller, Zaragosi,
  Barbry, Leslie, Kiem, Flynn, Fortune, Berger, Finberg, Kean, Garber, Schmidt,
  Lingwood, Shalek, Ordovas-Montanes, {HCA Lung Biological Network. Electronic
  address: lung-network@humancellatlas.org}, and {HCA Lung Biological
  Network}]{Ziegler2020-pb}
Carly G~K Ziegler, Samuel~J Allon, Sarah~K Nyquist, Ian~M Mbano, Vincent~N
  Miao, Constantine~N Tzouanas, Yuming Cao, Ashraf~S Yousif, Julia Bals,
  Blake~M Hauser, Jared Feldman, Christoph Muus, Marc~H Wadsworth, 2nd,
  Samuel~W Kazer, Travis~K Hughes, Benjamin Doran, G~James Gatter, Marko
  Vukovic, Faith Taliaferro, Benjamin~E Mead, Zhiru Guo, Jennifer~P Wang,
  Delphine Gras, Magali Plaisant, Meshal Ansari, Ilias Angelidis, Heiko Adler,
  Jennifer M~S Sucre, Chase~J Taylor, Brian Lin, Avinash Waghray, Vanessa
  Mitsialis, Daniel~F Dwyer, Kathleen~M Buchheit, Joshua~A Boyce, Nora~A
  Barrett, Tanya~M Laidlaw, Shaina~L Carroll, Lucrezia Colonna, Victor Tkachev,
  Christopher~W Peterson, Alison Yu, Hengqi~Betty Zheng, Hannah~P Gideon,
  Caylin~G Winchell, Philana~Ling Lin, Colin~D Bingle, Scott~B Snapper,
  Jonathan~A Kropski, Fabian~J Theis, Herbert~B Schiller, Laure-Emmanuelle
  Zaragosi, Pascal Barbry, Alasdair Leslie, Hans-Peter Kiem, Joanne~L Flynn,
  Sarah~M Fortune, Bonnie Berger, Robert~W Finberg, Leslie~S Kean, Manuel
  Garber, Aaron~G Schmidt, Daniel Lingwood, Alex~K Shalek, Jose
  Ordovas-Montanes, {HCA Lung Biological Network. Electronic address:
  lung-network@humancellatlas.org}, and {HCA Lung Biological Network}.
\newblock {SARS-CoV-2} receptor {ACE2} is an {Interferon-Stimulated} gene in
  human airway epithelial cells and is detected in specific cell subsets across
  tissues.
\newblock \emph{Cell}, 181\penalty0 (5):\penalty0 1016--1035.e19, May 2020.
\newblock ISSN 0092-8674, 1097-4172.
\newblock \doi{10.1016/j.cell.2020.04.035}.

\bibitem[Imoto et~al.(2002)Imoto, Goto, and Miyano]{Imoto2002-wm}
Seiya Imoto, Takao Goto, and Satoru Miyano.
\newblock Estimation of genetic networks and functional structures between
  genes by using bayesian networks and nonparametric regression.
\newblock \emph{Pac. Symp. Biocomput.}, pages 175--186, 2002.
\newblock ISSN 2335-6936, 2335-6928.

\bibitem[Kr{\"a}mer et~al.(2014)Kr{\"a}mer, Green, Pollard, and
  Tugendreich]{Kramer2014-mi}
Andreas Kr{\"a}mer, Jeff Green, Jack Pollard, Jr, and Stuart Tugendreich.
\newblock Causal analysis approaches in ingenuity pathway analysis.
\newblock \emph{Bioinformatics}, 30\penalty0 (4):\penalty0 523--530, February
  2014.
\newblock ISSN 1367-4803, 1367-4811.
\newblock \doi{10.1093/bioinformatics/btt703}.

\bibitem[Shannon et~al.(2003)Shannon, Markiel, Ozier, Baliga, Wang, Ramage,
  Amin, Schwikowski, and Ideker]{Shannon2003-va}
Paul Shannon, Andrew Markiel, Owen Ozier, Nitin~S Baliga, Jonathan~T Wang,
  Daniel Ramage, Nada Amin, Benno Schwikowski, and Trey Ideker.
\newblock Cytoscape: a software environment for integrated models of
  biomolecular interaction networks.
\newblock \emph{Genome Res.}, 13\penalty0 (11):\penalty0 2498--2504, November
  2003.
\newblock ISSN 1088-9051.
\newblock \doi{10.1101/gr.1239303}.

\end{thebibliography}


\captionsetup*{format=largeformat}

\renewcommand{\figurename}{\scriptsize Supplementary Fig.}
\setcounter{figure}{0}

\makeatletter
\renewcommand{\thefigure}{S\@arabic\c@figure}

\begin{figure*}[p]
\begin{center}
\includegraphics[width=.8\linewidth]{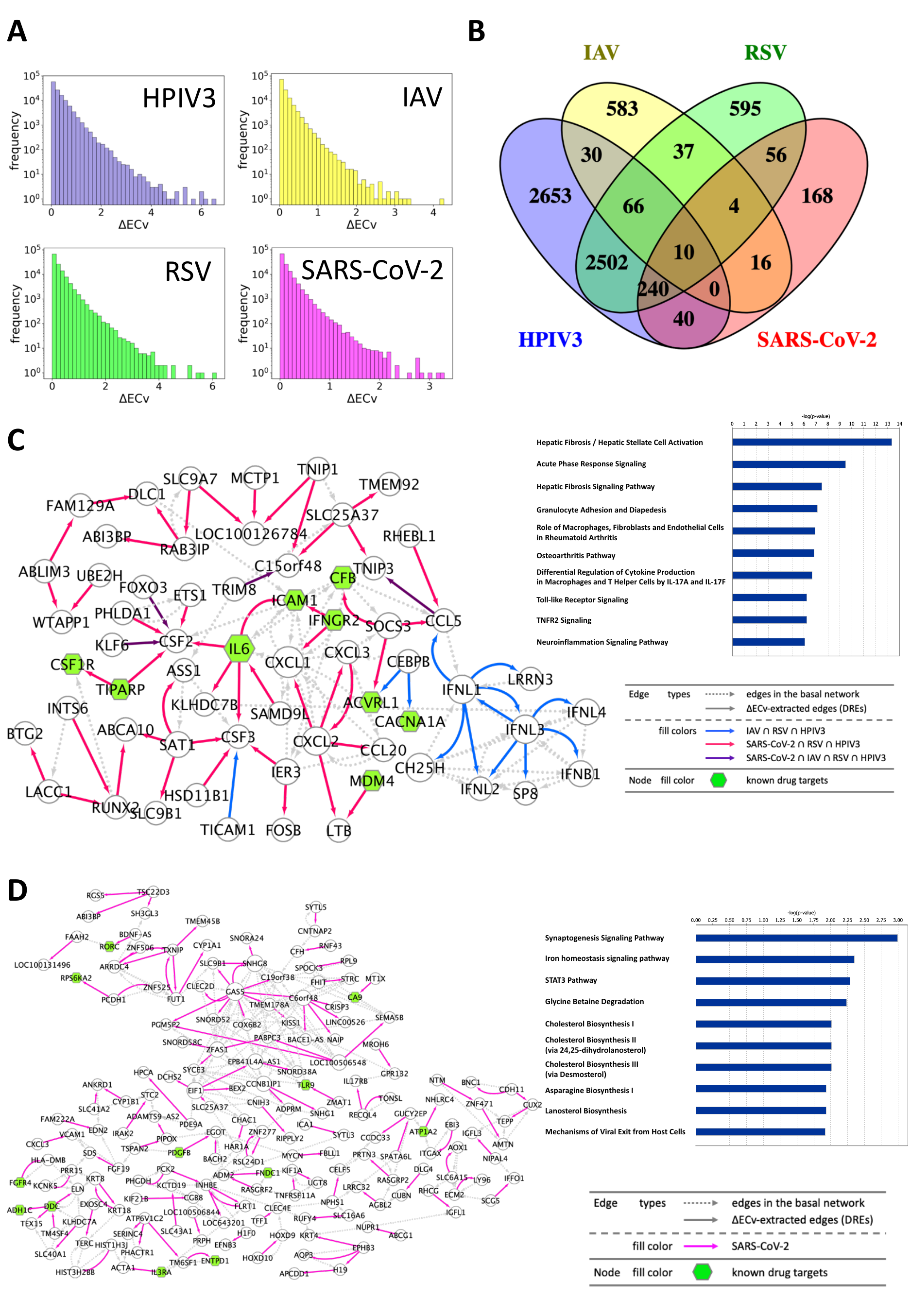}
\end{center}
\caption{\scriptsize Network comparison analyses across four respiratory viruses. (\textbf{A}) The histograms of $\Delta$ECv for each respiratory virus as indicated: SARS-CoV-2 (MOI: 2), HPIV3, IAV, and RSV. $\Delta$ECvs were calculated following the Eq. (2) where $S$ = virus-infected and $T$ = corresponding mock samples for each virus (see Methods). The X-axis corresponds to the threshold for each $\Delta$ECv. The Y-axis stands for the number of edges on a log scale. (\textbf{B}) The Venn diagram represents the numbers of $\Delta$ECv-extracted edges for all respiratory viruses with a $\Delta$ECv threshold of 1.0. (\textbf{C}) The respiratory viruses-shared network comprised of 62 nodes and 116 edges (including 53 basal edges). The colored solid edges represent DREs; SARS-CoV-2 $\cap$ IAV $\cap$ HPIV3 $\cap$ RSV (purple), SARS-CoV-2 $\cap$ RSV $\cap$ HPIV3 (red), IAV $\cap$ RSV $\cap$ HPIV3 (blue). The top 10 terms of canonical pathway analysis for the genes of $\Delta$ECv-extracted DREs shared by at least three viruses in the Venn diagram (Supplementary Fig. S1B). (\textbf{D}) SARS-CoV-2 specific network comprising 182 nodes and 295 edges (including 171 basal edges). The solid edges (magenta) represent DREs for SARS-CoV-2 (MOI: 2). The dot edges represent the basal edges (grey). The size of the node represents the extent of outdegree. The nodes (green) are target genes for existing drugs (Supplementary Table S1). The top 10 terms of canonical pathway analysis for the genes of $\Delta$ECv-extracted DREs exclusive for SARS-CoV-2 in the Venn diagram (Supplementary Fig. S1B).
}
\label{figS1}
\end{figure*}

\begin{figure*}[p]
\begin{center}
\includegraphics[width=.8\linewidth]{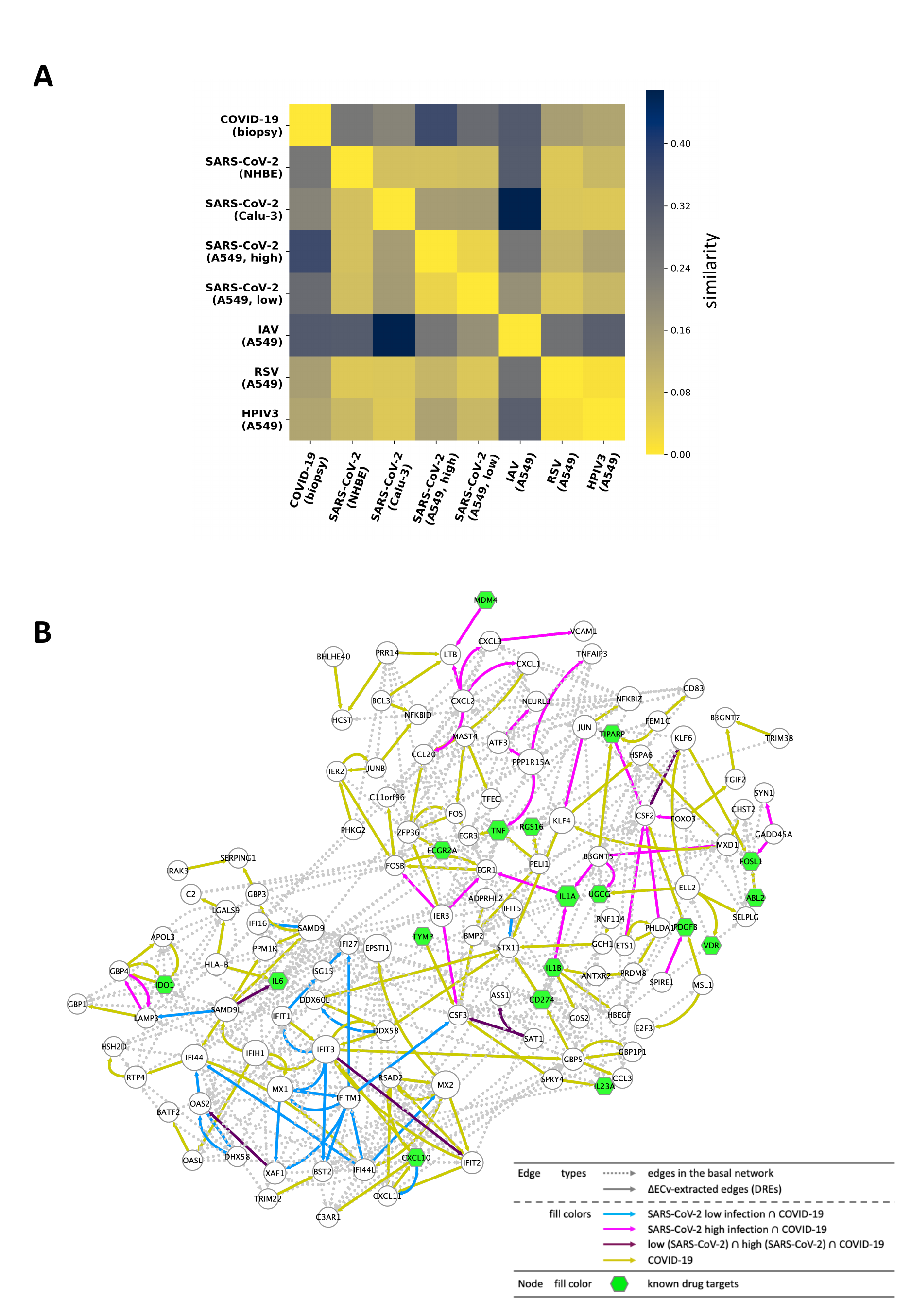}
\end{center}
\caption{\scriptsize Multiple analyses for generating the COVID-19-perturbated network.
(\textbf{A}) The similarity heatmap is shown and samples for comparisons are labeled as indicated. Similarity is calculated with cosine distance method for ECvs of the 167 DREs. (\textbf{B}) The COVID-19 patient-specific network in combination with the Venn diagram analysis (Fig. 5D). The network is composed of 127 nodes and 412 edges (including 245 basal edges). The colored solid edges represent DREs; SARS-CoV-2 (high MOI: 2) $\cap$ COVID-19-perturbated (magenta), SARS-CoV-2 (low MOI: 0.2) $\cap$  COVID-19-perturbated (blue), SARS-CoV-2 (high MOI: 2) $\cap$ SARS-CoV-2 (low MOI: 0.2) $\cap$ COVID-19-perturbated (purple), COVID-19-perturbated exclusive edges (yellow). The dot edges represent the basal edges (grey). The nodes (green) represent the known drug target genes (Supplementary Table S1). The node size represents the extent of outdegree.
}
\label{figS2}
\end{figure*}

\begin{figure*}[p]
\begin{center}
\includegraphics[width=\linewidth]{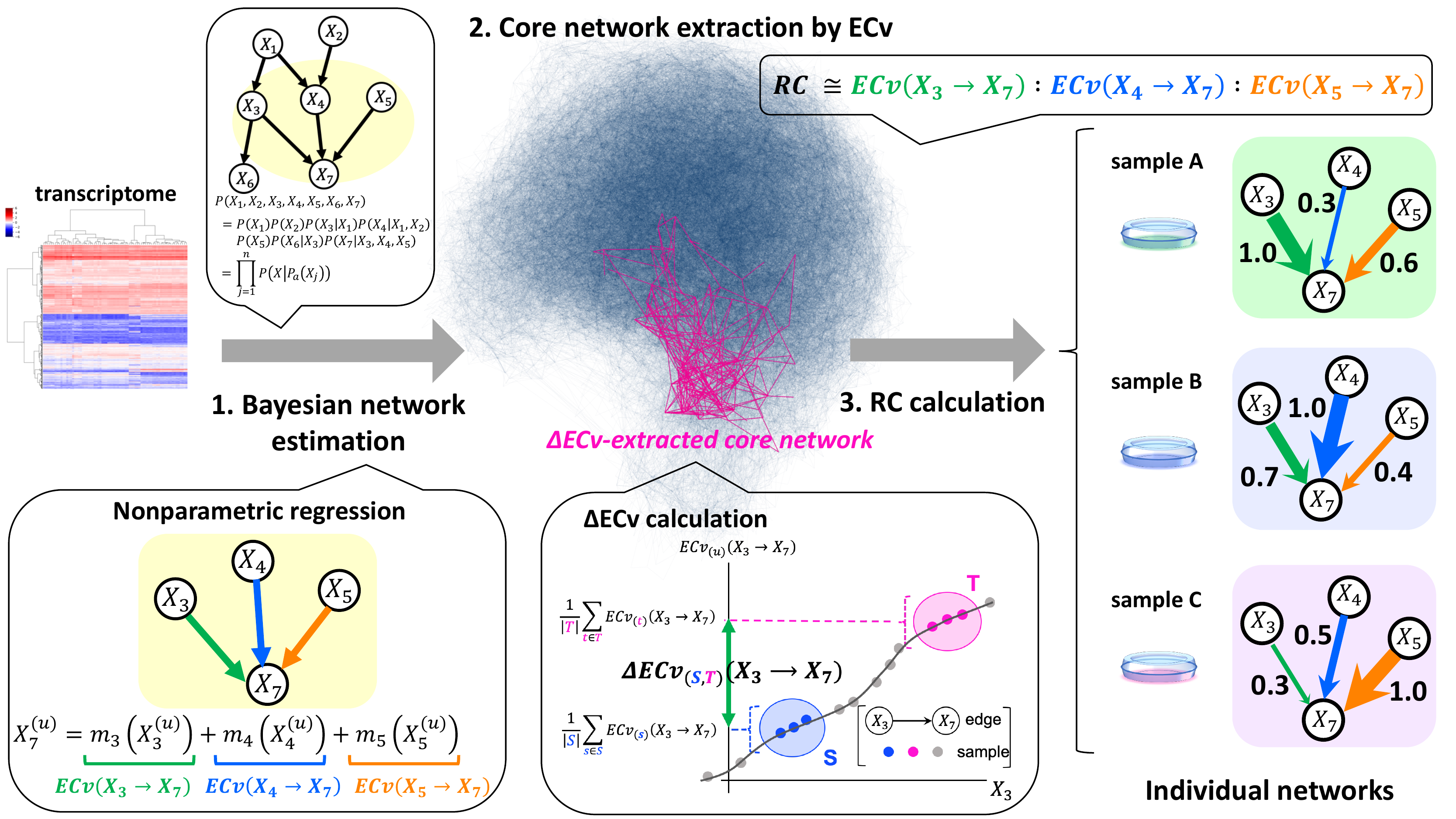}
\end{center}
\caption{\scriptsize Mathematical illustration of our proposed framework for the gene network analysis.
The centered hairball (blue) represents a basal network. The network (magenta) is a core network extracted by the $\Delta$ECv calculation.
}
\label{figS3}
\end{figure*}

\end{document}